\documentclass[preprint2, twocolappendix]{aastex631}

\usepackage[T1]{fontenc}
\DeclareRobustCommand{\VAN}[3]{#2}
\let\VANthebibliography\thebibliography
\def\thebibliography{\DeclareRobustCommand{\VAN}[3]{##3}\VANthebibliography}
\usepackage{graphicx}	
\usepackage{amsmath}	

\newcommand \logmhost {\log M_{\rm host}}
\newcommand \logmnsc {\log M_{\rm NSC}}
\usepackage{threeparttable}

\begin{document}

\title{Galaxy mass dependence of metal-enrichment of nuclear star clusters}

\author{Wenhe Lyu}
\affiliation{School of Astronomy and Space Sciences, University of Science and Technology of China, Hefei, 230026, People's Republic of China}
\affiliation{CAS Key Laboratory for Research in Galaxies and Cosmology, Department of Astronomy, \\ University of Science and Technology of China, Hefei 230026, People's Republic of China}

\author{Hong-Xin Zhang}
\thanks{Corresponding author. E-mail: hzhang18@ustc.edu.cn}
\affiliation{School of Astronomy and Space Sciences, University of Science and Technology of China, Hefei, 230026, People's Republic of China}
\affiliation{CAS Key Laboratory for Research in Galaxies and Cosmology, Department of Astronomy, \\ University of Science and Technology of China, Hefei 230026, People's Republic of China}

\author{Sanjaya Paudel}
\affiliation{Department of Astronomy, Yonsei University, Seoul, 03722, Republic of Korea}
\affiliation{Center for Galaxy Evolution Research, Yonsei University, Seoul, 03722, Republic of Korea}

\author{Tie Li}
\affiliation{School of Astronomy and Space Sciences, University of Science and Technology of China, Hefei, 230026, People's Republic of China}
\affiliation{CAS Key Laboratory for Research in Galaxies and Cosmology, Department of Astronomy, \\ University of Science and Technology of China, Hefei 230026, People's Republic of China}

\author{Yimeng Tang}
\affiliation{Department of Astronomy and Astrophysics, University of California Santa Cruz, 1156 High Street, Santa Cruz, CA 95064, USA}

\author{Guangwen Chen}
\affiliation{Sub-department of Astrophysics, University of Oxford, Keble Road, Oxford, OX1 3RH, UK}

\author{Xu Kong}
\affiliation{School of Astronomy and Space Sciences, University of Science and Technology of China, Hefei, 230026, People's Republic of China}
\affiliation{Deep Space Exploration Laboratory/Department of Astronomy, University of Science and Technology of China, Hefei, 230026, People's Republic of China}
\affiliation{Frontiers Science Center for Planetary Exploration and Emerging Technologies, University of Science and Technology of China, \\ Hefei, Anhui, 230026, People's Republic of China}

\author{Eric W. Peng}
\affiliation{NSF's NOIRLab, 950 N. Cherry Avenue, Tucson, AZ 85719, USA}



\date{Accepted XXX. Received YYY; in original form ZZZ}

\begin{abstract}

Nuclear Star Clusters (NSCs) are commonly found in galaxy centers, but their dominant formation mechanisms remain elusive. We perform a consistent analysis of stellar populations of 97 nearby NSCs, based on VLT spectroscopic data. The sample covers a galaxy stellar mass range of 10$^{7}$ to 10$^{11}$ M$_{\odot}$ and is more than 3$\times$ larger than any previous studies. We identify three galaxy stellar mass regimes with distinct NSC properties. In the low-mass regime of $\logmhost$ $\lesssim$ 8.5, nearly all NSCs have metallicities lower than circum-NSC host but similar to typical red globular clusters (GCs), supporting the GC inspiral-merger scenario of NSC formation. In the high-mass regime of $\logmhost$ $\gtrsim$ 9.5, nearly all NSCs have higher metallicities than circum-NSC host and red GCs, suggesting significant contributions from in-situ star formation (SF). In the intermediate-mass regime, a comparable fraction of NSCs have higher or lower metallicities than circum-NSC host and red GCs, with no clear dependence on NSC mass, suggesting intermittent in-situ SF. The majority of NSCs with higher metallicities than their host exhibit a negative age$-$metallicity correlation, providing clear evidence of long-term chemical enrichment. The average NSC$-$host metallicity difference peaks broadly around $\logmhost \sim 9.8$ and declines towards both higher and lower galaxy mass. We find that the efficiency of dynamical friction-driven inspiral of GCs observed in present-day galaxies can explain the NSC mass at $\logmhost \lesssim 9.5$ but falls short of observed ones at higher galaxy mass, reinforcing our conclusions based on stellar population analysis. 

\end{abstract}



\section{Introduction} 

Nuclear Star Clusters (NSCs) are dense and massive star clusters located near the center of galaxies. The mass and size of NSCs span a range that partially overlaps with that of globular clusters (GCs) at the low end, and reaches up to several tens of parsecs in half-light radius $R_{e}$ and several hundred million solar mass at the high end \citep[e.g.][]{Côté2006,Georgiev2016,Sánchez-Janssen2019,Neumayer2020}. The highest stellar densities in the Universe are found in NSCs. NSCs may serve as favorable formation sites of the elusive intermediate-mass black holes \citep[IMBH; e.g.,][]{Miller2009,Fragione2020,Atallah2023}. Moreover, tidally stripped NSCs are considered as an important contributor to the still mysterious ultra-compact dwarf galaxies \citep[UCD; e.g.,][]{Zhang2015,Liu2015,Ahn2017,zhang2018,Liu2020,Mayes2021,Chen2022,Wang2023}.

The formation mechanism of NSCs and its connection with host galaxies are not well understood. Statistical studies of the fraction of galaxies that host NSCs (i.e. nucleation fraction) in various environments \citep[e.g.,][]{Côté2006,Turner2012,denBrok2014, Sánchez-Janssen2019, Hoyer2021} suggest that the nucleation fraction has a strong dependence on galaxy stellar mass and reaches a maximum of $ \sim90\% $ in galaxies with $M_{\star}$ $\sim$ $ 10^{9}-10^{9.5} M_{\odot} $ and drops steadily towards both the lower and higher mass end.

The mass of NSCs (if exists) correlates positively with the stellar mass of the host galaxies, albeit with a substantial scatter. The correlation appears to be sub-linear, suggesting that lower mass galaxies tend to have a higher fraction of mass in NSCs than do higher mass galaxies \citep[e.g.,][]{Sánchez-Janssen2019,Neumayer2020,Hoyer2023a}. There is extensive evidence that NSCs may co-exist with central (super) massive black holes in many galaxies of stellar masses above 10$^{9}M_{\odot}$ \citep[see][ and references therein]{Neumayer2012, Neumayer2020}. The drop of nucleation fraction towards the higher mass end may be attributed to the influence of massive black holes \citep[e.g.,][]{Sedda2014, Antonini2015}.

The two commonly cited growth mechanisms for NSCs are GC inspiral and merger driven by dynamical friction \citep[e.g.,][]{Tremaine1975,Capuzzo-Dolcetta1993,Lotz2001, Sedda2014} and in-situ star formation triggered by torque induced gas inflow \citep[e.g.,][]{Silk1987,Mihos1994,Bekki2015,Feldmeier-Krause2015}. The dynamical friction-driven star cluster inspiral scenario is expected to be an unavoidable physical process, and is usually used to explain the presence of metal poor stellar populations found especially in the NSCs of dwarf galaxies \citep[]{Alfaro-Cuello2020, Fahrion2020, Fahrion2021}. It should be cautious that the surviving GC population can be quite different from that spiraled into the galactic center.
In the latter mechanism, NSCs are expected to have relatively extended star formation histories and contain chemically enriched young stellar populations.
The two distinct mechanisms are not necessarily mutually exclusive in reality. \cite{Guillard2016} bring forward a hybrid mechanism, whereby gas-bearing young massive star clusters spiral into galaxy center due to dynamical friction and continue forming stars after migrating to the center.

Dynamical friction-driven inspiral of massive GCs is probably unavoidable from a theoretical point of view, but observational verification is largely indirect \citep[e.g.,][]{Neumayer2020}. In particular, there appears to be a central deficit of massive GCs in many early-type galaxies \citep[e.g.,][]{Lotz2001}. The nucleation fraction seems to track the fraction of galaxies hosting GCs for low-mass early-type galaxies \citep[][]{Sánchez-Janssen2019}. Recently, \cite{Román2023} found an unusually high concentration of GC candidates near the center of UGC7346, and speculated that this may be a NSC caught in its early stage of formation through dynamical friction-driven inspiral of GCs \cite[see also][for a similar discovery]{Schiavi2021}.
On the other hand, the in-situ formation scenario has a wealth of observational evidence \citep[e.g.,][]{Walcher2006,Seth2006,Kacharov2018,Fahrion2021}, with the NSC in our Milky Way being a notable example \citep[e.g.,][]{Feldmeier-Krause2015}.

Recent literature seems to reach a broad consensus that the dominant formation pathway has a strong connection with galaxy mass \citep[][and references therein]{Côté2006, Sedda2014, Neumayer2020}, in the sense that NSCs grow primarily by dynamical friction-driven GC infall and merger at galaxy stellar mass much smaller than $10^{9}M_{\odot}$, while above the transition mass of $\sim 10^{9}M_{\odot}$, in-situ star formation gradually becomes the dominant pathway. At the transition galaxy mass of $\sim 10^{9}M_{\odot}$ (the typical NSC mass being $\sim10^{6}M_{\odot} $), both GC inspiral and in-situ star formation play important roles. Such a galaxy mass-dependent formation pathways appears to be supported by several observational findings, such as galaxy mass dependent NSC stellar populations (this work), NSC shapes \citep[e.g.,][]{Spengler2017} and nucleation fractions. It is not clear, however, what drives this apparent mass-dependent dichotomy \citep[][]{Neumayer2020}

Stellar population properties provide important clues to the dominant formation pathways of NSCs. With an exception of the NSC in the Milky Way, where individual stars are well resolved, integrated-light spectroscopy is usually the practical choice for obtaining robust stellar population estimation of extragalactic NSCs \citep[e.g.,][]{Rossa2006,Walcher2006,Seth2006,Paudel2011,Kacharov2018,Johnston2020,Fahrion2021,Fahrion2022b}. A general finding from these spectroscopic analysis is that NSCs in low-mass galaxies (<$10^{9}M_{\odot}$) tend to have light-weighted metallicities lower than their host galaxies, whereas NSCs in more massive galaxies tend to have higher stellar metallicities than their host galaxies. This is in line with the expectation of the galaxy mass-dependent dominant growth pathways mentioned above. However, the existing spectroscopic analysis of NSCs is limited to relatively small sample sizes, which makes it difficult to draw robust conclusions about the dependence of NSC stellar populations on host galaxy properties. So far, stellar population analysis of the largest samples of NSCs were performed by \cite{Paudel2011} (26 nucleated dwarf elliptical galaxies in the Virgo cluster) and \cite{Fahrion2021} (25 nucleated early-type galaxies mostly in the Fornax cluster) respectively. 

Besides the relatively small sample size of previous spectroscopic analysis of stellar populations of NSCs, different studies are usually different in their adopted spectral extraction methods, stellar population models, or spectral modeling techniques. The potential systematic bias induced by these differences hinder a statistical analysis of literature samples based on a direct compilation of stellar population parameters derived from different studies. In this work, we collect high-quality archival optical spectra of nearby NSCs observed by various instruments on the Very Large Telescope (VLT), and perform a statistical analysis of the stellar population properties obtained through consistent spectral extraction and modeling.

The paper is structured as follows. A description of the sample and data reduction is given in Sec~\ref{sec:Data and reduction} and the analysis method is given in Sec~\ref{sec:analysis}. Results are presented in Sec~\ref{sec:result}. The discussion and summary are given in Sections~\ref{sec:discussion} and \ref{sec:conclusion}, respectively.

\section{Sample and data reduction}
\label{sec:Data and reduction}
\subsection{Sample and observations}
Most of the spectroscopic observations of extra-galactic NSCs have been carried out with VLT \citep[also HST, see e.g.][]{Rossa2006}. We start with the catalog of nucleated galaxies in the Local Volume \citep{Karachentsev2013}, Virgo Cluster and Fornax Cluster, as compiled by \cite{Hoyer2021}, and search the ESO Science Archive for optical spectroscopic observations of the nuclear regions of these galaxies. The search returns 97 galaxies in total, among which 12 were observed with the X-Shooter spectrograph, 40 with the FORS2 spectrograph, and 45 with the MUSE spectrograph. We note that 29 of these NSCs do not have stellar population analysis based on spectral modeling before. We download and calibrate the raw data of the 97 nucleated galaxies. A brief description of the sample and data observed by each instrument is given in Table \ref{tab:detail}.

\begin{table*}
\caption{Detail of nucleated galaxies. Col(1):Name of galaxy; Col(2),(3):Ra, Dec (J2000) (From NED); Col(4):Stellar mass of host galaxy, refer to \protect\cite{Fahrion2021}, \protect\cite{Fahrion2022a}, \protect\cite{Hoyer2021}, \protect\cite{Consolandi2016}, \protect\cite{Pechetti2020}, \protect\cite{Relatores2019}; Col(5):Stellar mass of NSC, refer to \protect\cite{Fahrion2021}, \protect\cite{Fahrion2022a}, \protect\cite{Kacharov2018}, \protect\cite{Spengler2017}, \protect\cite{W.Graham2009}, \protect\cite{Hoyer2023a}, \protect\cite{Hoyer2023b}, \protect\cite{Nguyen2018}, \protect\cite{Nguyen2022}, \protect\cite{Sánchez-Janssen2019}, \protect\cite{Fahrion2020}, \protect\cite{Carlsten2022}, \protect\cite{Pechetti2020}, \protect\cite{Georgiev2016}, \protect\cite{Calzetti2015}; Col(6):Hubble morphological type \protect\citep{DeVaucouleurs1991}, with negative numbers assigned to early-type galaxies and positive numbers to late-type galaxies, refer to \protect\cite{Kourkchi2017}; Col(7):Instruments; Col(8):Source paper of these nucleated galaxies (P11:\protect\cite{Paudel2011}, K18:\protect\cite{Kacharov2018}, J20:\protect\cite{Johnston2020}, F20:\protect\cite{Fahrion2020}, F21:\protect\cite{Fahrion2021}, F22:\protect\cite{Fahrion2022b}, Un:Unpublish for NSCs research). Note:(*) Due to lack of literature data, stellar mass of NSCs in NGC1705, NGC2784, NGC3368, NGC3489, VCC0592, VCC0765, VCC0786 and VCC0871 are estimated from $(\rm M/L)_{V}$ and integral V band luminosity after correction of flux loss in slit.
(**) We also use data from programma 096.B-0063, 097.B-0761 and 098.B-0239 for FCC207 and 1100.B-0651 for NGC2835.}
\hspace{-3.4cm}
\begin{threeparttable}
\label{tab:detail}
\setlength{\tabcolsep}{1.5mm}{
\begin{tabular}{ccccccccc}
\hline
(1) & (2) & (3) & (4) & (5) & (6) & (7) & (8) & (9)\\
Name & RA(J2000) & DEC(J2000) & $\log(M_{\rm host})$ & $\log(M_{\rm NSC})$ & Hubble Type & Instrument & Source & PID\\
- & degree & degree & $M_{\odot}$ & $M_{\odot}$ & - & - & - & -\\
\hline
NGC247 & 11.786 & $-$20.760 & 9.258 & 6.25 & 6.9 & X-Shooter & K18 & 084.B-0499\\
NGC2784 & 138.081 & $-$24.173 & 10.672 & $ 8.64^{*} $ & $-$2.1 & X-Shooter & Un & 097.B-0435\\
NGC300 & 13.723 & $-$37.684 & 9.129 & 5.99 & 6.9 & X-Shooter & K18 & 084.B-0499\\
NGC3115 & 151.308 & $-$7.719 & 10.866 & 7.18 & $-$2.9 & X-Shooter & Un & 097.B-0435\\
NGC3621 & 169.569 & $-$32.814 & 9.739 & 7.0 & 6.9 & X-Shooter & K18 & 086.B-0651\\
NGC5068 & 199.728 & $-$21.039 & 9.479 & 6.22 & 6.0 & X-Shooter & Un & 097.B-0435\\
NGC5102 & 200.490 & $-$36.630 & 9.32 & 7.86 & $-$2.8 & X-Shooter & K18 & 086.B-0651\\
NGC5206 & 203.433 & $-$48.151 & 9.361 & 7.18 & $-$2.9 & X-Shooter & K18 & 084.B-0499\\
NGC5236 & 204.254 & $-$29.865 & 10.579 & 7.38 & 5.0 & X-Shooter & Un & 097.B-0435\\
NGC628 & 24.174 & 15.784 & 10.223 & 7.06 & 5.2 & X-Shooter & Un & 098.B-0024\\
NGC7713 & 354.062 & $-$37.938 & 8.808 & 5.61 & 6.7 & X-Shooter & Un & 097.B-0435\\
NGC7793 & 359.458 & $-$32.591 & 9.36 & 6.96 & 7.4 & X-Shooter & K18 & 084.B-0499\\
VCC0216 & 184.255 & 9.408 & 8.56 & 6.18 & $-$2.5 & FORS2 & P11 & 078.B-0178\\
VCC0308 & 184.712 & 7.862 & 9.112 & 6.49 & $-$5.0 & FORS2 & P11 & 078.B-0178\\
VCC0389 & 185.014 & 14.962 & 9.491 & 7.02 & $-$3.6 & FORS2 & P11 & 078.B-0178\\
VCC0490 & 185.412 & 15.745 & 9.100 & 6.76 & $-$0.2 & FORS2 & P11 & 078.B-0178\\
VCC0545 & 185.582 & 15.734 & 7.556 & 6.73 & $-$2.2 & FORS2 & P11 & 078.B-0178\\
VCC0592 & 185.712 & 13.593 & 7.909 & $ 6.05^{*} $ & $-$5.0 & FORS2 & Un & 085.B-0971\\
VCC0725 & 186.101 & 15.075 & 7.33 & 6.02 & $-$5.0 & FORS2 & P11 & 078.B-0178\\
VCC0765 & 186.265 & 13.245 & 7.666 & $ 6.17^{*} $ & $-$5.0 & FORS2 & Un & 085.B-0971\\
VCC0786 & 186.310 & 11.850 & 9.17 & $ 6.95^{*} $ & $-$5.0 & FORS2 & Un & 085.B-0971\\
VCC0856 & 186.491 & 10.054 & 9.092 & 6.77 & $-$5.0 & FORS2 & P11 & 078.B-0178\\
VCC0871 & 186.524 & 12.560 & 7.467 & $ 5.54^{*} $ & $-$5.0 & FORS2 & Un & 085.B-0971\\
VCC0916 & 186.638 & 12.743 & 8.876 & 6.50 & $-$5.0 & FORS2 & Un & 085.B-0971\\
VCC0929 & 186.669 & 8.436 & 7.405 & 7.28 & $-$0.8 & FORS2 & P11 & 078.B-0178\\
VCC0940 & 186.696 & 12.454 & 9.143 & 6.7 & $-$5.0 & FORS2 & Un & 085.B-0971\\
VCC0965 & 186.763 & 12.561 & 8.565 & 6.6 & $-$5.0 & FORS2 & Un & 085.B-0971\\
VCC0990 & 186.821 & 16.024 & 9.186 & 6.83 & $-$3.7 & FORS2 & P11 & 078.B-0178\\
\hline
\end{tabular}}
\end{threeparttable}
\vspace{-27.74pt}
\end{table*}

\begin{table*}
\hspace{-3.4cm}
\begin{threeparttable}
\label{tab:detail2}
\setlength{\tabcolsep}{1.5mm}{
\begin{tabular}{ccccccccc}
\hline
Name & RA(J2000) & DEC(J2000) & $\log(M_{\rm host})$ & $\log(M_{\rm NSC})$ & Hubble Type & Instrument & Source & PID\\
- & degree & degree & $M_{\odot}$ & $M_{\odot}$ & - & - & - & -\\
\hline
VCC1069 & 187.027 & 12.898 & 8.222 & 6.1 & $-$5.0 & FORS2 & Un & 085.B-0971\\
VCC1073 & 187.036 & 12.093 & 9.21 & 6.7 & $-$4.2 & FORS2 & Un & 085.B-0971\\
VCC1104 & 187.117 & 12.824 & 8.659 & 6.2 & $-$5.0 & FORS2 & Un & 085.B-0971\\
VCC1122 & 187.174 & 12.916 & 8.78 & 6.5 & $-$1.7 & FORS2 & Un & 085.B-0971\\
VCC1167 & 187.311 & 7.878 & 8.72 & 7.03 & $-$5.0 & FORS2 & P11 & 078.B-0178\\
VCC1185 & 187.348 & 12.451 & 8.424 & 6.19 & $-$5.0 & FORS2 & P11 & 078.B-0178\\
VCC1254 & 187.521 & 8.073 & 8.012 & 7.04 & $-$5.0 & FORS2 & P11 & 078.B-0178\\
VCC1261 & 187.543 & 10.779 & 9.64 & 6.66 & $-$4.8 & FORS2 & P11 & 078.B-0178\\
VCC1304 & 187.666 & 15.130 & 8.67 & 6.93 & $-$2.2 & FORS2 & P11 & 078.B-0178\\
VCC1308 & 187.691 & 11.343 & 8.232 & 6.35 & $-$5.0 & FORS2 & P11 & 078.B-0178\\
VCC1333 & 187.755 & 7.723 & 8.04 & 6.7 & $-$5.0 & FORS2 & P11 & 078.B-0178\\
VCC1348 & 187.816 & 12.332 & 8.836 & 7.33 & $-$5.0 & FORS2 & P11 & 078.B-0178\\
VCC1353 & 187.831 & 12.738 & 8.02 & 6.4 & $-$4.4 & FORS2 & P11 & 078.B-0178\\
VCC1355 & 187.834 & 14.115 & 8.931 & 6.08 & $-$5.0 & FORS2 & P11 & 078.B-0178\\
VCC1386 & 187.964 & 12.657 & 9.096 & 6.50 & $-$4.9 & FORS2 & Un & 085.B-0971\\
VCC1389 & 187.967 & 12.482 & 8.125 & 6.49 & $-$5.0 & FORS2 & P11 & 078.B-0178\\
VCC1407 & 188.011 & 11.890 & 8.785 & 6.29 & $-$5.0 & FORS2 & P11 & 078.B-0178\\
VCC1431 & 188.097 & 11.263 & 9.204 & 6.63 & $-$4.7 & FORS2 & Un & 085.B-0971\\
VCC1491 & 188.308 & 12.858 & 8.823 & 5.9 & $-$3.7 & FORS2 & Un & 085.B-0971\\
VCC1661 & 189.103 & 10.385 & 8.36 & 6.47 & $-$5.0 & FORS2 & P11 & 078.B-0178\\
VCC1826 & 190.047 & 9.896 & 8.584 & 6.59 & $-$5.0 & FORS2 & P11 & 078.B-0178\\
VCC1861 & 190.244 & 11.184 & 8.959 & 6.51 & $-$4.9 & FORS2 & P11 & 078.B-0178\\
VCC1945 & 190.725 & 11.438 & 8.73 & 6.67 & $-$2.2 & FORS2 & P11 & 078.B-0178\\
VCC2019 & 191.335 & 13.693 & 9.004 & 6.78 & $-$4.2 & FORS2 & P11 & 078.B-0178\\
CIRCINUS & 213.291 & $-$65.339 & 10.485 & 7.569 & 3.3 & MUSE & Un & 0103.B-0396\\
ESO59-01 & 112.826 & $-$68.188 & 7.896 & 6.16 & 9.8 & MUSE & F22 & 0108.B-0904\\
FCC119 & 53.391 & $-$33.573 & 9.047 & 6.81 & $-$2.9 & MUSE & F21 & 296.B-5054\\
FCC148 & 53.820 & $-$35.266 & 9.883 & 8.37 & $-$2.2 & MUSE & F21 & 296.B-5054\\
FCC153 & 53.879 & $-$34.447 & 9.88 & 7.29 & $-$2.1 & MUSE & F21 & 296.B-5054\\
FCC170 & 54.132 & $-$35.295 & 10.543 & 8.45 & $-$2.1 & MUSE & F21 & 296.B-5054\\
FCC177 & 54.198 & $-$34.740 & 9.992 & 7.83 & $-$1.9 & MUSE & F21 & 296.B-5054\\
FCC182 & 54.226 & $-$35.375 & 9.505 & 6.03 & $-$2.8 & MUSE & F21 & 296.B-5054\\
FCC188 & 54.269 & $-$35.591 & 8.89 & 6.85 & $-$4.3 & MUSE & F21 & 096.B-0399\\
FCC190 & 54.287 & $-$35.195 & 9.995 & 7.18 & $-$2.8 & MUSE & F21 & 296.B-5054\\
FCC193 & 54.299 & $-$35.746 & 10.501 & 8.15 & $-$2.8 & MUSE & F21 & 296.B-5054\\
FCC202 & 54.527 & $-$35.440 & 9.137 & 6.76 & $-$3.0 & MUSE & F21 & 094.B-0895\\
$\rm FCC207^{**}$ & 54.580 & $-$35.129 & 8.68 & 6.06 & $-$4.3 & MUSE & J20 & 094.B-0576\\
\hline
\end{tabular}}
\end{threeparttable}
\end{table*}

\begin{table*}
\hspace{-3.4cm}
\begin{threeparttable}
\label{tab:detail3}
\setlength{\tabcolsep}{1.5mm}{
\begin{tabular}{ccccccccc}
\hline
Name & RA(J2000) & DEC(J2000) & $\log(M_{\rm host})$ & $\log(M_{\rm NSC})$ & Hubble Type & Instrument & Source & PID\\
- & degree & degree & $M_{\odot}$ & $M_{\odot}$ & - & - & - & -\\
\hline
FCC211 & 54.590 & $-$35.259 & 8.942 & 6.7 & $-$4.5 & MUSE & F21 & 096.B-0399\\
FCC215 & 54.657 & $-$35.758 & 6.79 & 5.94 & $-$3.5 & MUSE & F21 & 096.B-0399\\
FCC222 & 54.805 & $-$35.371 & 7.431 & 6.45 & $-$2.2 & MUSE & F21 & 096.B-0399\\
FCC223 & 54.831 & $-$35.726 & 8.78 & 6.38 & $-$4.9 & MUSE & F21 & 096.B-0399\\
FCC227 & 54.959 & $-$35.523 & 6.73 & 6.06 & $-$3.5 & MUSE & F21 & 096.B-0399\\
FCC245 & 55.141 & $-$35.023 & 8.77 & 6.05 & $-$4.3 & MUSE & F21 & 101.C-0329\\
FCC249 & 55.175 & $-$37.511 & 10.106 & 6.93 & $-$4.9 & MUSE & F21 & 296.B-5054\\
FCC255 & 55.265 & $-$33.779 & 9.648 & 6.98 & $-$2.1 & MUSE & F21 & 296.B-5054\\
FCC277 & 55.595 & $-$35.154 & 9.786 & 7.22 & $-$2.9 & MUSE & F21 & 296.B-5054\\
FCC301 & 56.265 & $-$35.973 & 9.662 & 6.91 & $-$3.3 & MUSE & F21 & 296.B-5054\\
FCC306 & 56.439 & $-$36.347 & 7.426 & 6.13 & 7.9 & MUSE & J20 & 296.B-5054\\
FCC310 & 56.557 & $-$36.696 & 10.01 & 7.81 & $-$1.9 & MUSE & F21 & 296.B-5054\\
FCC47 & 51.634 & $-$35.714 & 9.926 & 8.74 & $-$3.0 & MUSE & F21 & 060.A-9192\\
FCCB1241 & 54.569 & $-$35.508 & 8.13 & 5.48 & $-$5.0 & MUSE & F21 & 102.B-0455\\
IC1959 & 53.302 & $-$50.414 & 7.926 & 6.13 & 8.5 & MUSE & F22 & 0108.B-0904\\
IC5332 & 353.615 & $-$36.101 & 9.866 & 6.842 & 6.8 & MUSE & Un & 1100.B-0651\\
KK197 & 200.508 & $-$42.535 & 6.993 & 6.04 & 10.0 & MUSE & F20 & 0101.A-0193\\
KKs58 & 206.503 & $-$36.329 & 6.875 & 5.87 & $-$3.7 & MUSE & F20 & 0101.A-0193\\ 
NGC1487 & 58.942 & $-$42.368 & 8.95 & 6.01 & 7.2 & MUSE & F22 & 0100.B-0116\\
NGC1705 & 73.557 & $-$53.361 & 8.207 & $ 6.75^{*} $ & $-$2.7 & MUSE & Un & 094.B-0745\\
NGC1796 & 75.677 & $-$61.140 & 9.060 & 6.82 & 5.3 & MUSE & F22 & 0108.B-0904\\
$\rm NGC2835^{**}$ & 139.470 & $-$22.355 & 9.528 & 6.61 & 5.0 & MUSE & Un & 098.B-0551\\
NGC3274 & 158.071 & 27.669 & 8.472 & 5.632 & 6.7 & MUSE & Un & 0110.B-0125\\
NGC3368 & 161.691 & 11.820 & 10.506 & $ 7.89^{*} $ & 2.1 & MUSE & Un & 0104.B-404\\
NGC3489 & 165.077 & 13.901 & 10.27 & $ 7.83^{*} $ & $-$1.2 & MUSE & Un & 0104.B-404\\
NGC3593 & 168.654 & 12.818 & 10.27 & 7.22 & $-$0.4 & MUSE & Un & 0106.B-0359\\
NGC4592 & 189.828 & $-$0.532 & 9.02 & 5.8 & 8.0 & MUSE & F22 & 095.B-0532\\
NGC5253 & 204.983 & $-$31.640 & 8.645 & 5.512 & 8.9 & MUSE & Un & 094.B-0745\\
NGC853 & 32.922 & $-$9.306 & 9.3 & 6.305 & 8.7 & MUSE & F22 & 0108.B-0904\\
UGC3755 & 108.466 & 10.522 & 7.75 & 4.777 & 9.9 & MUSE & F22 & 0108.B-0904\\
UGC5889 & 161.843 & 14.069 & 7.881 & 6.09 & 8.9 & MUSE & F22 & 0108.B-0904\\
UGC8041 & 193.803 & 0.117 & 8.83 & 6.74 & 6.9 & MUSE & F22 & 0104.D-0503\\
\hline
\end{tabular}}
\end{threeparttable}
\end{table*}

\subsubsection{X-Shooter spectra}
X-shooter is a multi-wavelength, medium resolution long-slit spectrograph mounted at the UT3 Cassegrain focus of VLT \citep{Vernet2011}. It covers the entire ultraviolet to near-infrared wavelength range simultaneously with three spectroscopic arms: UVB (2980-5600\AA), VIS (5500-10200\AA) and NIR (10200-24800\AA).

The X-Shooter spectra of 12 nucleated galaxies were taken through three observing programs (084.B-0499(C), 086.B-0651(C) and 097.B-0435(B)). Spectral modeling for 6 of the 12 NSCs was presented in \cite{Kacharov2018}. These observations were carried out with a 11\arcsec~slit length for all the spectroscopic arms. The slit widths are 0.8\arcsec, 0.7\arcsec~and 0.6\arcsec~respectively for the UVB, VIS, and NIR arms. In this work, only the UVB (R$\sim$6700) and VIS (R$\sim$11400) data are used for analysis. The raw data are reduced with the ESO REFLEX X-Shooter pipeline (v.3.5.3), which performs bias subtraction, flat field correction, order tracing, wavelength calibration, flux calibration, flexure compensation, sky subtraction with dedicated offset sky exposures, image combination, etc. The final products are rectified 2-D spectra and error maps.

\subsubsection{FORS2 spectra}
The FORS2 is a multi-mode optical instrument mounted at UT1 of VLT. Among the galaxies with available FORS2 spectra, 26 were observed through the multi-object mode and 14 through the long-slit mode. All the spectra were taken with a V300 grism, a wavelength coverage of 3300 \AA-11000 \AA, a slit width of 1\arcsec, with a corresponding spectral resolution of $\sim$ 11 \AA~full width half maximum (at 5000\AA). \cite{Paudel2011} presented spectral analysis for the 26 NSCs observed with the multi-object mode. The 14 long-slit spectra were acquired with a 40\arcsec~slit length. The reader is referred to \cite{Paudel2010} for more details about the observational layout and data reduction.

\subsubsection{MUSE spectra}
MUSE is an integral-field spectrograph mounted at UT4 of VLT. The MUSE data used here were taken in the wide field mode of MUSE, which provides a 1\arcmin $\times$ 1\arcmin~field of view (FoV) and a wavelength coverage of 4700-9300\AA. The spectra have a spectral resolution of $\sim$ 2.5 \AA~full width half maximum (at 7000\AA), with a small wavelength dependence \citep{Bacon2017}. The spatial sampling is 0.2\arcsec $\times$ 0.2\arcsec~and the spectral sampling is 1.25\AA~per pixel. The raw data are reduced with the ESO REFLEX MUSE pipeline (v.2.8.7), following the standard procedures. To keep our analysis of the MUSE spectra consistent with that of the long-slit spectra from X-Shooter and FORS2, we make a artificial slit across NSC and extract a subcube of the central 1.2\arcsec~$\times$~40\arcsec region of the reduced MUSE data cube for each galaxy.\ The following analysis of MUSE data will be based on these subcubes.

\subsection{Spectral extraction of NSCs and their host}
Spectra of the nuclear region of our galaxies are mainly from the NSCs, but have non-negligible contribution from the underlying host stellar populations. Therefore, it is necessary to perform a subtraction of the underlying host light for a clean analysis of the NSCs. To this end, we follow a procedure similar to that of \cite{Paudel2011}. Specifically, we take the following approach to obtain NSC spectra clean of host contamination.

At the typical distance of our galaxies, the NSCs are unresolved or at most marginally resolved by the seeing-limited ground-based observations. Therefore, the spatial profile near the NSCs may be described by a superposition of a seeing-defined Gaussian profile (representing the NSCs) and a S\'ersic profile (representing the underlying host). To perform the profile decomposition, we first collapse the 2-D long slit spectra along the wavelength direction to obtain a white light spatial profile. Then, the white light spatial profile is fitted with a combination of a Gaussian and S\'ersic functions. The Gaussian function is defined by the standard deviation $\sigma$ and peak flux, and the S\'ersic function is defined by S\'ersic index $n$, effective radius ($R_{e}$), and central surface brightness. After obtaining the best-fit parameters, the relative flux contribution of the S\'ersic host to the nuclear region can be quantified. Our profile decomposition suggests that the local flux ratio of the Gaussian component to the S\'ersic component falls well below 0.1 at radius $R$ > 4$\sigma$. Therefore, we consider the region beyond 4$\sigma$ (and < 6$\sigma$) to be representative of the underlying host of NSCs, and restrict the spectral extraction of NSCs to $R < 2\sigma$ region. Particularly, a scaling factor $ C $ is first calculated by using the best-fit S\'ersic component:
\begin{equation}
C=\frac{\int_{0}^{2\sigma}f_{\rm Sersic} \, dR}
{\int_{4\sigma}^{6\sigma}f_{\rm Sersic} \, dR}
\end{equation}
Then, NSC spectra that are clean of host contamination can be obtained as follows:
\begin{equation}	f_{\lambda}^{\rm NSC}={\int_{0}^{2\sigma}f_{\lambda,R}^{\rm obs} \, dR}-C\times{\int_{4\sigma}^{6\sigma}f_{\lambda,R}^{\rm obs} \, dR}
\end{equation}
where $f_{\lambda,R}^{\rm obs}$ is the observed flux density at given wavelength $\lambda$ and radius. As described above, the spectra of the representative host underlying NSCs can be simply derived as:
\begin{equation}	f_{\lambda}^{\rm host}={\int_{4\sigma}^{6\sigma}f_{\lambda,R}^{\rm obs} \, dR}
\end{equation}
Note that, in the above calculation, we have assumed that the scale factor $ C $ does not vary with wavelength. A wavelength dependent solution for $C$ is preferable, but it would be subject to significant uncertainties in practice, due to a lower S/N.


Lastly, all of the NSC spectra and host spectra are corrected for the Galactic extinction by using the python packages \textsc{sfdmap} \citep{Schlegel1998} and \textsc{extinction} \citep{Barbary2016}, assuming the \cite{Cardelli1989} extinction law with $R_{v}$$=3.1 $. 

\section{Analysis}
\label{sec:analysis}

\subsection{Full spectrum fitting}
\label{sec:ppxf}

We fitted all spectra by using full spectrum fitting technique with the python package Penalized PiXel-Fitting (pPXF) \citep{Cappellari2004,Cappellari2017}, which is a well developed tool for extracting stellar population properties and kinematics by fitting weighted combinations of single stellar population (SSP) models to an observed spectrum, without a prior assumption about the functional form of the star formation history. In this work, we adopt MILES SSP model spectra \citep{Vazdekis2010}, which is generated based on the MILES spectral library,  the BaSTi scaled-solar isochrones, and a Kroupa initial stellar mass function (IMF). The SSP models cover a range of ages from 0.03 Gyr to 14 Gyr and a range of metallicities [M/H] from $-$2.27  to +0.40. The model spectra have a spectral resolution of  2.51~\AA~in the wavelength range from 3525 to 7500~\AA. Our spectral fitting is performed over the wavelength range from 4000 to 6800~\AA~for the X-Shooter and FORS2 data, while from 4700 to 6800 {\AA} for the MUSE data. To keep the analysis consistent for different data sets, the X-Shooter and MUSE spectra (and the model spectra) are smoothed to the same spectral resolution of $ \sim11  $ {\AA} of the FORS2 data prior to the spectral modeling. We follow a procedure similar to \cite{Tang2022} to carry out the pPXF fitting.

(i) Firstly, we fit each spectrum for redshift and velocity dispersion by invoking additive and multiplicative polynomials of degree 10 to accommodate the continuum shape difference between models and observations. With the above fitting, the line-of-sight velocity and the velocity dispersion are determined and fixed in the subsequent fitting.

(ii) Secondly, with the kinematic parameters being fixed, we perform a second pPXF fitting to each spectrum without regularization, by invoking multiplicative polynomials to adjust the continuum of the models. The resultant polynomial parameters define a correction curve that accounts for flux calibration inaccuracy and internal dust reddening.\ In this round of fitting, the noise spectrum is re-scaled such that the minimum reduced $\chi^{2}_{\rm r}$ is equal to 1. The correction curve, re-scaled noise spectrum, and residual spectrum between the best-fit model and data are used in the following fitting.

(iii) Thirdly, we fit for stellar age and [M/H] via a wild bootstrapping method \citep{DAVIDSON2008}, where we add generated noise(-noise or noise with probability=0.5) to the observed spectrum according to the residual spectrum derived above for 100 times, and perform pPXF fitting to each of the noise-disturbed spectra by invoking a mild regularization (regul=10). The 100 fittings result in distributions for best-fit light- and mass-weighted ages and [M/H]. These distributions are used determine the most probable values of ages, [M/H] and their associated 68\% confidence intervals. The results for our sample are presented in Table ~\ref{tab:ppxf result1}.

\subsection{[\texorpdfstring{$\alpha$}{}/Fe] estimation} \label{sec:alpha2fe_estimate}
\begin{figure}
	\includegraphics[width=\columnwidth]{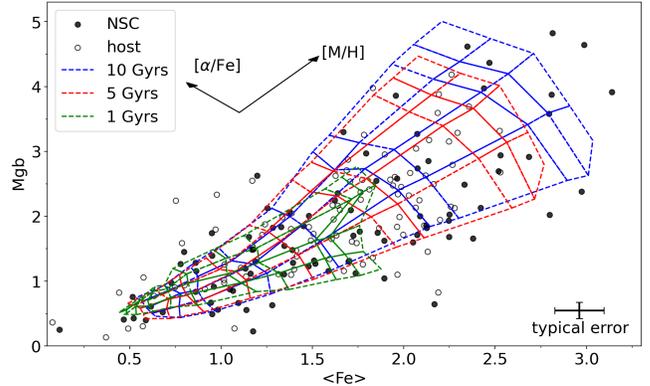}
	\caption{The Mg$b$-<Fe> index-index diagram. Measurements of both the NSCs (filled circles) and underlying host (hollow circles) of our sample are plotted. The MILES SSP model grids for three different stellar ages (10 Gyr in blue, 5 Gyr in red and 1 Gyr in green) are overlaid for illustration. In practice, the [$\alpha$/Fe] parameter is estimated by interpolating the model grids for given age(as obtained from the full-spectrum fitting). For the small number of measurements falling outside the model grids, the [$\alpha$/Fe] is estimated through a linear extrapolation of the model grids.}
	\label{fig:alphaFe}
\end{figure}
$\alpha$ elements are primarily produced in short lived high-mass stars ($\gtrsim$ 8 M$_{\odot}$) and are released via core-collapse supernovae, whereas for the Fe-peak elements, both core-collapse supernovae and thermonuclear supernovae have significant contributions (thermonuclear supernovae have main contributions). Therefore, the relatively longer delay times of thermonuclear supernovae ($\gtrsim$ 1 Gyr) than the core-collapsed supernovae results in a higher [$\alpha$/Fe] ratio for stellar systems formed earlier or over shorter timescales \citep[e.g.][]{Thomas2003}.

To estimate [$\alpha$/Fe] for our sample, we follow \cite{Thomas2003} and take the Lick absorption-line index Mg$b$ as a proxy for $\alpha$ elements and the composite Fe Lick index <Fe> (= (Fe5270+Fe5335)/2) as a proxy for Fe-peak elements. We adopt the semi-empirical MILES stellar population models with variable [$\alpha$/Fe] \citep{Knowles2023}, which provides a uniform coverage of [$\alpha$/Fe] from $-$0.2 to +0.6. As with our full-spectrum fitting, we choose the Kroupa IMF and smooth the model spectra to a resolution of $11 $ {\AA}. To measure the Lick indices (Mg$b$, Fe5270, Fe5335) on both the model and observed spectra, we use python-based package \textsc{pyphot} \citep{Fouesneau2022}. There is a weak but non-negligible dependence of the Mg$b$/<Fe> ratio on stellar age and metallicity. Therefore, for each NSC or host spectrum, we only consider models that match the light-weighted age obtained from our full-spectrum fitting, and obtain the [$\alpha$/Fe] values through interpolation of the model grids with respect to the [$\alpha$/Fe] parameter. For a small number of index measurements that fall outside the model grids, we estimate the [$\alpha$/Fe] values through a linear extrapolation. The uncertainties of [$\alpha$/Fe] are determined by randomly disturbing the measured index values according to the measurement uncertainties and repeating the above procedure of [$\alpha$/Fe] estimation.

The distribution of the NSCs and circum-NSC host on the Mg$b$ vs. <Fe> diagram is shown in Figure~\ref{fig:alphaFe}, where the model grids of different [$\alpha$/Fe] at three representative ages are overlaid for illustration. In Figure~\ref{fig:alphaFe}, the small filled circles and hollow circles represent the measured indices of NSCs and underlying hosts, respectively. A small number of spectra (12 NSCs and 3 hosts) have negative values for one of the two Fe Lick indices, which may be attributed to remarkable noise disturbance. In this case, only the Fe index with positive measurement is used in the estimation of [$\alpha$/Fe] based on model grids involving only a single Fe Lick index. Lastly, we note that the spectra of 10 galaxies are corrupted near wavelength range that defines the Mg$b$ or both of the two Fe indices, so we ignore the 10 galaxies whenever the [$\alpha$/Fe] parameter is relevant in the following analysis.

\subsection{Comparison with measurements in the literature}
\begin{figure}
  \centering
  \includegraphics[width=\columnwidth]{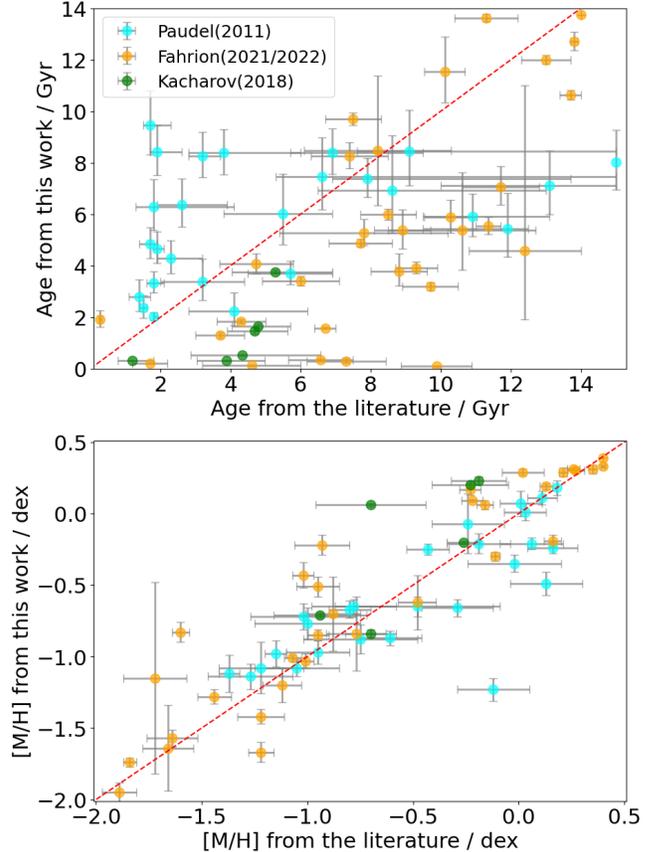}
  \caption{Comparison of ages (top) and [M/H] (bottom) of the NSC from the literature and this work. The red dashed lines refer to the one-to-one relation.}
  \label{fig:agemhcompare}
\end{figure}

Here we compare our light-weighted result of age and [M/H] with the result in the literature which are from \cite{Paudel2011},  \cite{Kacharov2018}, \cite{Fahrion2021} and \cite{Fahrion2022a}. The comparisons for age and [M/H] are respectively shown in the upper and lower panels of Figure~\ref{fig:agemhcompare}. It is worth noting that the spectral extraction and stellar population modeling in the literature are significantly different from ours. Particularly, \cite{Paudel2011} estimated the stellar population parameters based on Lick indices, instead of full-spectrum fitting as adopted in our work. \cite{Kacharov2018} and \cite{Fahrion2022a} did not subtract the flux contamination of the underlying hosts in the NSCs spectra. 

In Figure~\ref{fig:agemhcompare}, we show that our [M/H] estimation are in reasonable agreement with the literature values. The results of age appear to have substantial difference for many NSCs, especially when compared to that based on Lick indices. This comparison may partially reflect the fact that [M/H] is better constrained by the integrated-light spectra than ages, and also demonstrate the potential problem of statistical analysis of stellar population properties based on a direct compilation of estimates from the literature.

\subsection{Sample classification based on the projected phase-space diagram} \label{sec:ppsdiagram}
\begin{figure}
    \centering    
    \includegraphics[width=\columnwidth]{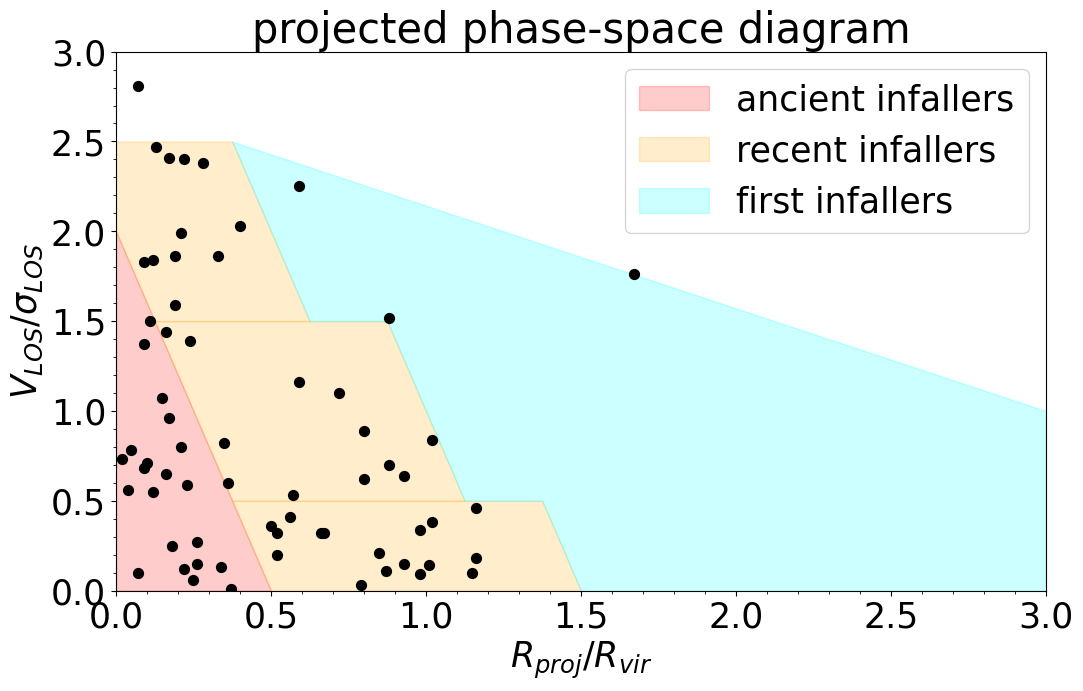}
    \caption{Projected phase-space diagram, being used to classify our galaxies into different enviroment. Each black dot is a galaxy associated with the Virgo or Fornax cluster in our sample.  x-axis: the projected distance of a galaxy from the cluster center normalized by the virial radius of the cluster. y-axis: the line-of-sight velocity difference between the galaxy and the cluster center normalized by the velocity dispersion of the cluster. The projected phase-space diagram is split into three color-coded regions, i.e., first infallers, recent infallers and ancient infallers, largely following \protect\cite{Rhee2017}. Note that galaxies in Local Volume are not shown here.}
    \label{fig:phase2}
\end{figure}

To explore the environmental dependence of NSC properties, we turn to the projected phase-space diagram. The projected phase-space diagram for a cluster of galaxies involves the projected distance to the cluster center and the line-of-sight velocity of individual galaxies relative to the cluster. Recent simulations \citep[e.g.][]{Rhee2017,Smith2019} suggest that galaxy location in the projected phase-space diagram is, in a statistical sense, an indicator of the infall time and thus the environmental effect on the evolution of the galaxies in question. Many of our galaxies are associated with the two nearest galaxy clusters: Virgo cluster and Fornax cluster. The distribution of our galaxies on the projected phase-space diagram is shown in Figure \ref{fig:phase2}, where the projected distance and line-of-sight velocities with respect to the cluster center have been normalized, respectively, by the virial radius and velocity dispersion of the host cluster.

The demarcation lines for phase-space regions occupied by galaxies with distinct infall times in a statistical sense from \cite{Rhee2017} are also plotted in Figure \ref{fig:phase2}. Following \cite{Rhee2017}, we classify the cluster galaxies in our sample broadly into three subsamples as indicated by the regions filled with different colors in Figure \ref{fig:phase2}, i.e., first infallers, recent infallers and ancient infallers, where the first infallers refer to galaxies that are falling toward the virial radius for the first time, the recent infallers refer to galaxies that have crossed the virial radius in the recent couple of Gyr, and the ancient infallers refer to galaxies that are close to be virialized in the cluster. In what follows, we will group the galaxies not associated with the two galaxy clusters in our sample into the above defined subsample of first infallers and redefine them as Local Volume or first infallers (LV or first infallers).


\section{Results}
\label{sec:result}

\subsection{Mass-metallicity relation}
\label{sec:MZR}
\begin{figure*}
  \centering
  \includegraphics[width=\textwidth]{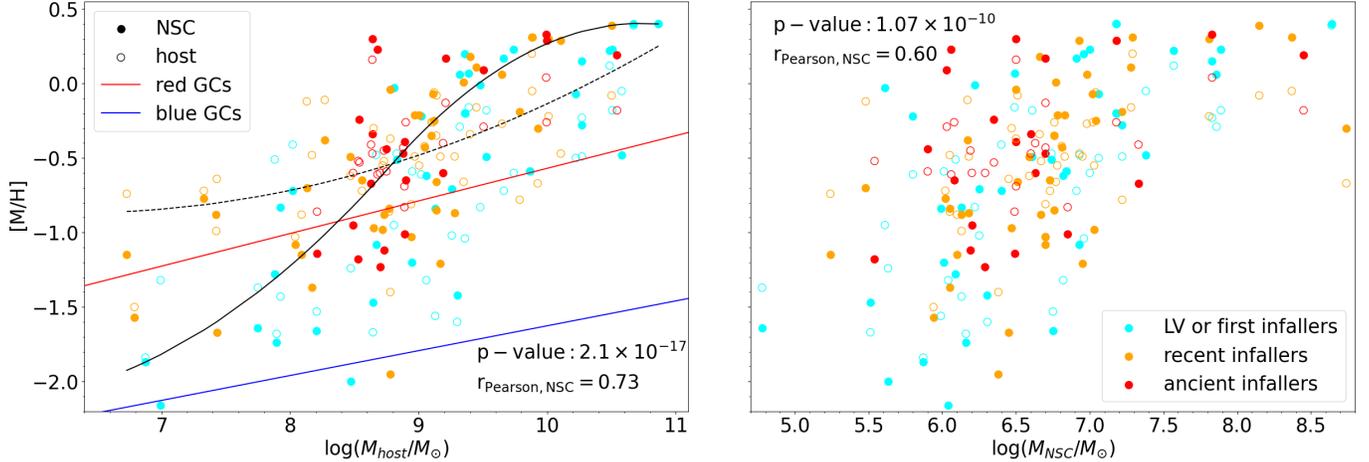}
  \caption{Mass-metallicity distribution of the NSCs and the host. The [M/H] of NSCs and the circum-NSC host is plotted against the galaxy stellar mass in the left panel and against the NSC stellar mass in the right panel. In both panels, the NSCs are represented as filled circles, while the host is represented as open circles. The data points are color coded according to the projected phase-space environment classification, as indicated in the right panel and illustrated in Figure \ref{fig:phase2}. The black solid line and black dashed line respectively represents a constrained B-splines smooth curve fitting for the 50\% (median) quantile of [M/H] of NSCs and the circum-NSC host. According to \protect\cite{Peng2006}, red line and blue line represent mean metallicities of red GC populations and blue GC populations as a function of galaxy stellar mass.} 
  \label{fig:massmetal}
\end{figure*}

The mass-metallicity distributions of our sample are presented in Figure~\ref{fig:massmetal}, where the left panel plots [M/H] of NSCs and their host as a function of the galaxy stellar mass and the right panel plots [M/H] as a function of the NSCs stellar mass.\ To guide the eye, a constrained B-splines smooth curve fitting for the 50\% (median) quantile of [M/H] distribution of NSCs (solid line) and circum-NSC host (dashed line) is performed by using the CRAN package ``cobs''. In addition, the linear relations between mean metallicities of GCs and the host galaxy stellar mass, as obtained by \cite{Peng2006}, are also overplotted in the left panel. We note that the \cite{Peng2006} relations were calibrated based on the metallicity scale of \cite{Zinn1984}, which, although being referred as [Fe/H], has been shown to trace the total metallicity rather than iron abundance \citep[][]{Thomas2003}.

The NSC metallicities have a significant correlation with their host galaxy mass, with the Pearson correlation coefficient 
$r$ = 0.73 (P-value = 2.1$\times$10$^{-17}$), whereas the NSC metallicities have a weaker correlation with the NSC mass ($r$ = 0.60). 
The NSCs have systematically lower average metallicities than the host at $\log M_{\rm host}$ $\lesssim$ 9, whereas the reverse is true at higher galaxy stellar masses. We will explore the metallicity difference between NSCs and their host in more detail in the next subsection.

In comparison to GCs, the metallicities of NSCs exhibit a steeper overall galaxy mass dependence, particularly at $\logmhost$ $\lesssim$ 9.5. \ Compared to the blue GCs, NSCs have systematically higher metallicities (with few exceptions) across the galaxy stellar mass range explored here. At $\log M_{\rm host}$ $\lesssim$ 8.5, NSCs have average metallicities slightly lower than typical red GCs for given galaxy stellar mass, while at $\log M_{\rm host}$ $>$ 8.5, NSCs have increasingly higher metallicities than typical red GCs towards the higher galaxy mass end.

Based on the above results, we can already infer that inspiral-merger of classical GCs cannot be an important formation channel for NSCs residing in galaxies of $\log M_{\rm host}$ $\gtrsim$ 9, confirming previous claims based on much smaller samples and inhomogeneous estimation. In contrast, typical NSCs in galaxies with $\log M_{\rm host}$ $\lesssim$ 8.5 were probably formed at similar or even early epoch with their typical red GCs, prior to the formation of the bulk of their circum-NSC host stellar populations. Therefore, GC inspiral-merger is probably the dominant formation chanel for NSCs in galaxies of $\log M_{\rm host}$ $\lesssim$ 8.5. 
In a similar vein, NSCs in galaxies of intermediate stellar masses may have a mixed formation mechanisms, involving both GC inspiral-merger and in-situ star formation.

\subsection{The metallicity difference between NSCs and their host} \label{sec:metaldiff}

\begin{figure*}
  \centering
  \includegraphics[width=\textwidth]{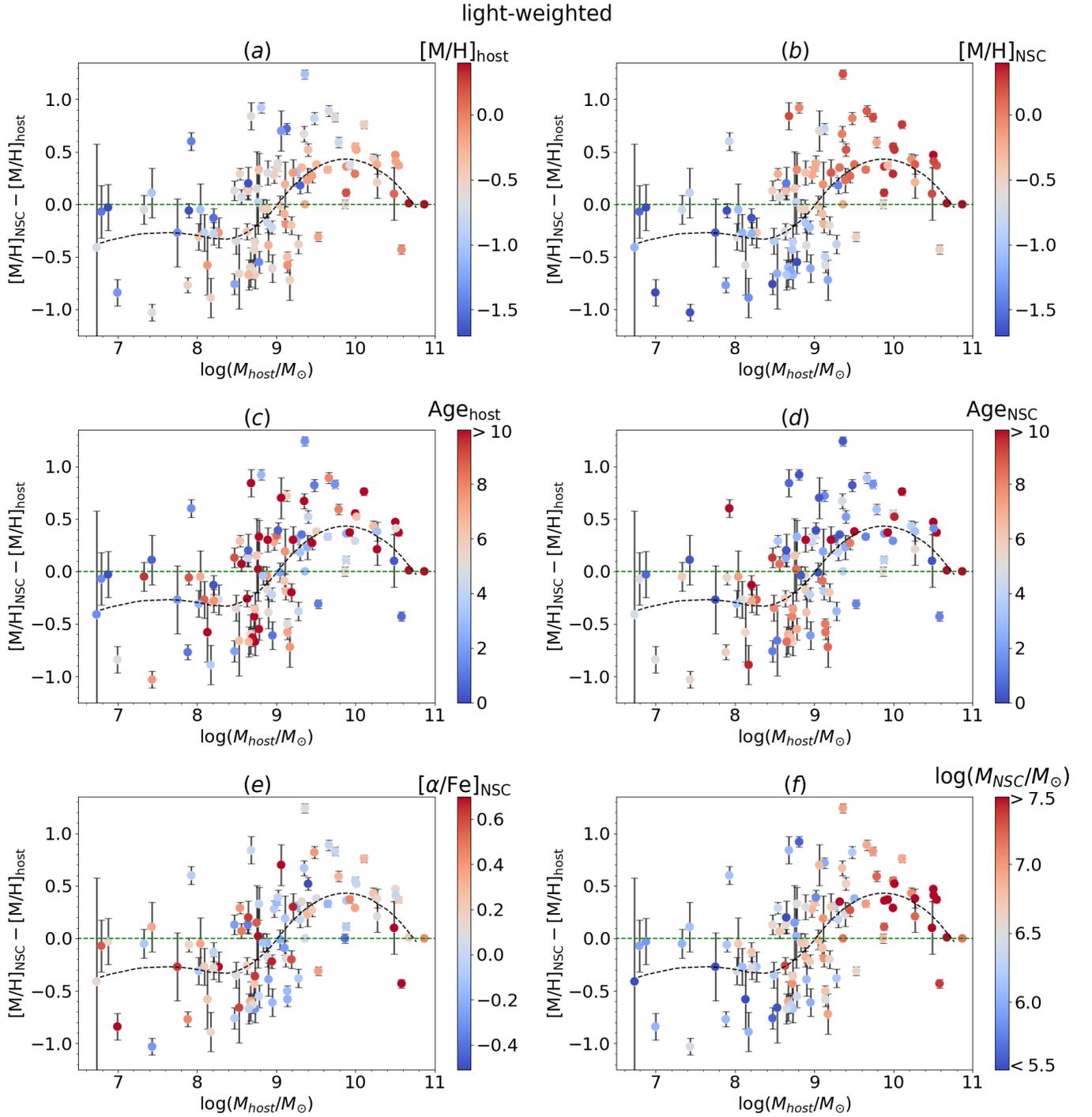}
  \caption{
   Metallicity difference of NSCs and circum-NSC host ([M/H]$_{\rm NSC}$-[M/H]$_{\rm host}$) vs. host galaxy stellar mass. The data points are color-coded by different stellar population properties in different subplots, as indicated in the color bar titles. The green dashed horizontal line in each subplot marks a zero metallicity difference. In each subplot, a constrained B-splines smooth curve fitting for the 50\%(median) quantile of [M/H]$_{\rm NSC}$-[M/H]$_{\rm host}$ is represented by a black dashed curve.}
  \label{fig:metaldiff}
\end{figure*}

As is evident in Figure \ref{fig:massmetal}, typical GCs have metallicities lower than the circum-NSC host across the galaxy stellar mass range explored here, so NSCs  owning lower metallicities than the underlying host are expected to form through inspiral-merger of metal-poor star clusters. On the other hand, if the growth of NSCs is sustained by continuous gas inflow from large galactocentric radii, and the inflowing gas is expected to be progressively metal-enriched as it flows through the disk or halo. So NSCs formed in-situ over extended timescales may have comparable or even higher metallicities than the circum-NSC host.

In Figure~\ref{fig:metaldiff}, the metallicity differences of [M/H]$_{\rm NSC}$-[M/H]$_{\rm host}$ (hereafter [M/H]$_{\rm diff}$) are plotted against the galaxy stellar mass, where the data points in different panels are color-coded by various light-weighted stellar population properties, including metallicity of host galaxies, metallicity of NSCs, age of NSCs, age of the circum-NSC host, [$\alpha$/Fe] of NSCs and stellar mass of NSCs.

Casting attention to the overall trend of [M/H]$_{\rm diff}$ vs. $M_{\rm host}$, there is virtually no galaxy mass dependence of [M/H]$_{\rm diff}$ at $\logmhost$ $\lesssim$ 8.5, where the average [M/H]$_{\rm diff}$ is $\sim$ $-$0.3 with a substantial scatter. Nearly all galaxies with $\logmhost$ $\lesssim$ 8.5 have [M/H]$_{\rm diff}$ < 0. In contrast, at the high galaxy mass end of $\logmhost$ $\gtrsim$ 9.5, nearly all galaxies have [M/H]$_{\rm diff}$ $\geq$ 0. Galaxies in the intermediate mass range of 8.5 $-$ 9.5 have average [M/H]$_{\rm diff}$ $\sim$ 0.0, with a comparable number of galaxies above or below the average. Moreover, the average [M/H]$_{\rm diff}$ appears to follow a unimodal distribution as a function of $\logmhost$, with a peak of average [M/H]$_{\rm diff}$ $\sim$ 0.5 near $\logmhost$ $\sim$ 9.8. The average [M/H]$_{\rm diff}$ drops steadily towards both the lower (until reaching below $\logmhost$ $\sim$ 8.5) and higher mass end, reaching [M/H]$_{\rm diff}$ $\sim$ 0 at $\logmhost$ $\sim$ 11. We note that the peak of the average [M/H]$_{\rm diff}$ distribution is at a higher galaxy mass than does the well-known unimodal galaxy mass dependence of nucleation fraction, which peaks near $\logmhost$ $\sim$ 9.0 \citep[e.g.][]{Hoyer2021} where the average [M/H]$_{\rm diff}$ $\sim$ 0. We also note that the large metallicity difference between NSCs and the circum-NSC host can not be explained by the well-established negative radial metallicity gradients of galaxies, which predict [M/H]$_{\rm diff}$ values well below 0.1 dex.

To understand the drivers of the spread of [M/H]$_{\rm diff}$, we explore various stellar population properties in different subplots of Figure \ref{fig:metaldiff}. By comparing subplots a and b, we find that galaxies with positive [M/H]$_{\rm diff}$ have systematically more metal-enriched NSCs than galaxies with negative [M/H]$_{\rm diff}$. Looking more closely at the galaxy mass dependence of [M/H]$_{\rm diff}$, the large spread of [M/H]$_{\rm diff}$ at the intermediate galaxy mass (8.5 $\lesssim$ $\logmhost$ $\lesssim$ 9.5) is primarily attributed to the spread of [M/H]$_{\rm NSC}$, whereby galaxies with larger [M/H]$_{\rm diff}$ tend to have larger [M/H]$_{\rm NSC}$ rather than smaller [M/H]$_{\rm host}$. In contrast to the intermediate mass galaxies, the [M/H]$_{\rm diff}$ spread at $\logmhost$ $\gtrsim$ 9.5 is primarily attributed to the spread of [M/H]$_{\rm host}$ rather than [M/H]$_{\rm NSC}$, whereby galaxies with larger [M/H]$_{\rm diff}$ tend to have lower [M/H]$_{\rm host}$, with a remarkable scatter. In addition, stellar age is not closely correlated with the spread of [M/H]$_{\rm diff}$, except that most NSCs (80\%) with [M/H]$_{\rm diff}$ > 0.5 have relatively young NSC ages ($\leq$ 5 Gyr; subplot d). Lastly, neither [$\alpha$/Fe] (subplot e) nor $M_{\rm NSC}$ (subplot f) is correlated with the [M/H]$_{\rm diff}$ spread at given galaxy stellar mass.

The above findings suggest that the connection between metal enrichment of NSCs and their underlying host is dependent on galaxy stellar mass. Particularly, only for the intermediate-mass galaxies, the metal-enrichment level of NSCs largely drives the spread of [M/H]$_{\rm diff}$. In contrast, in the high galaxy mass regime, the metal-enrichment level of NSCs appears to be largely "saturated", albeit with substantial scatter, and galaxies with larger [M/H]$_{\rm diff}$ tend to have lower [M/H]$_{\rm host}$, which is probably attributed to low-metallicity gas inflow toward the circum-NSC host region. In addition, it is intriguing that more metal-enriched NSCs in the intermediate-mass galaxies tend not to have larger NSC mass. This implies a lack of synchrony between metal enrichment and mass growth of NSCs.

\subsection{Age-metallicity distributions}
\label{sec:agemetal}
\begin{figure*}
  \centering
  \includegraphics[width=\textwidth]{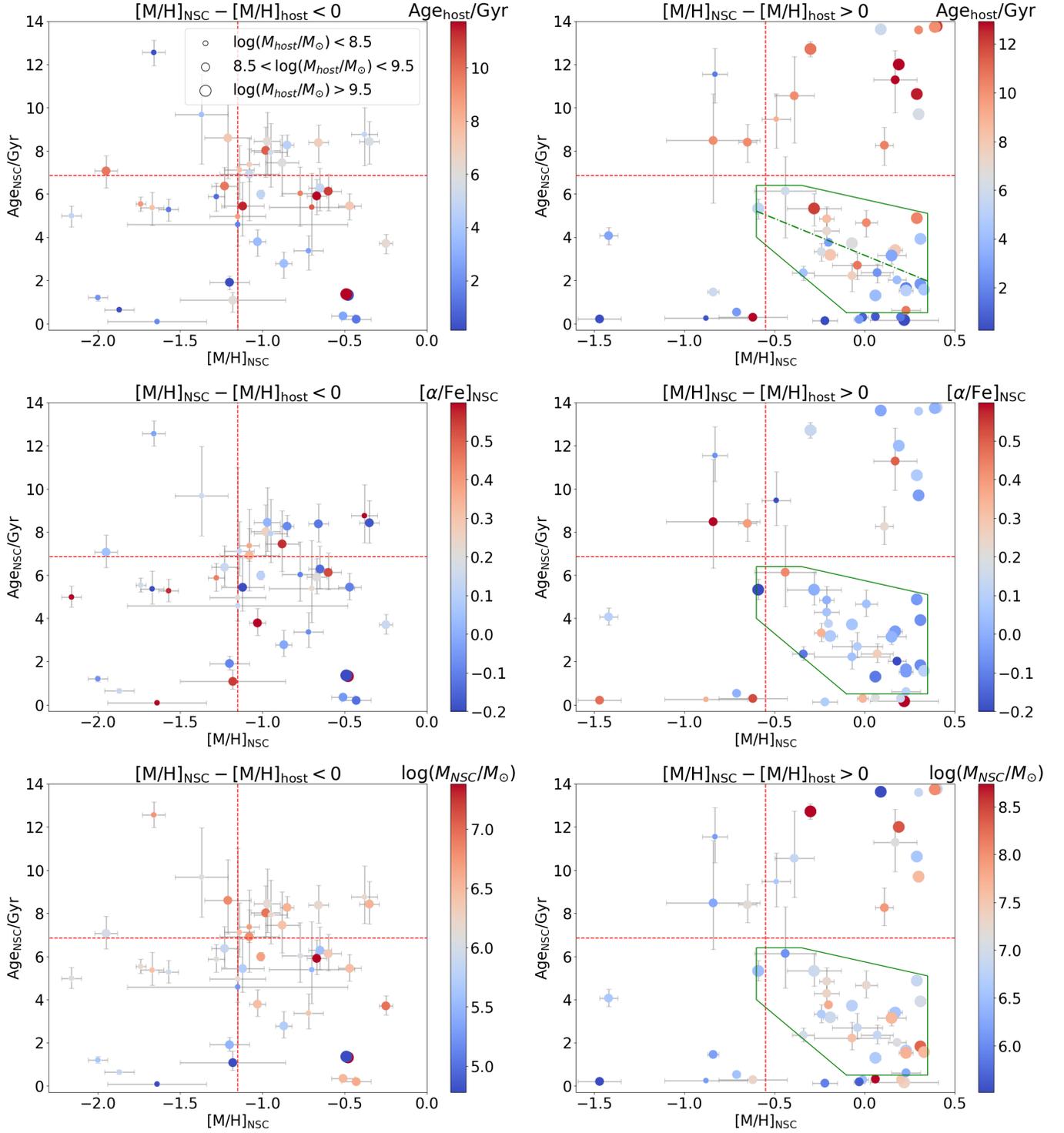}
  \caption{Age$-$metallicity distribution of NSCs with [M/H]$_{\rm NSC}$ < [M/H]$_{\rm host}$ (left panels) and [M/H]$_{\rm NSC}$ > [M/H]$_{\rm host}$ (right panels). The data points are color-coded according to the light-weighted Age$_{\rm host}$ (first row), [$\alpha$/Fe]$_{\rm NSC}$ (middle row), and $\logmnsc$ (bottom row). The symbol sizes are in accord with the host galaxy stellar mass ranges, as indicated in the legend. The green solid polygon in the right panels encloses data points belonging to the same GMM group, and the green dash-dotted line is the best-fit linear relation to NSCs enclosed by the green polygon. See Section \ref{sec:agemetal} for details.}
  \label{fig:agemetal-coloraam}
\end{figure*}

Age-metallicity relation is a powerful probe of the evolutionary history of stellar systems. The age-[M/H] distributions of our sample are presented in Figure \ref{fig:agemetal-coloraam}, where the galaxies/NSCs are distinguished according to  [M/H]$_{\rm diff}$, galaxy stellar mass, age of the circum-NSC host,  [$\alpha$/Fe] of the NSCs and $\logmnsc$.

The subsamples of NSCs with [M/H]$_{\rm diff}$ < 0 and > 0 have distinct age-[M/H] distributions. Specifically, a majority of NSCs with [M/H]$_{\rm diff}$ < 0 are clustered within a narrow age interval centering near $\sim$ 7 Gyr and a narrow [M/H]$_{\rm NSC}$ interval between $-$1.1 and $-$0.5. Except for this clustering trend, there is no obvious age-[M/H] correlation for NSCs with [M/H]$_{\rm diff}$ < 0. In contrast, NSCs with [M/H]$_{\rm diff}$ > 0 appear to consist of two major groups in the age-[M/H] diagram, whereby one group of NSCs ($\sim$ 25\%) randomly occupy the old-age ($\gtrsim$ 7 Gyr) and high-[M/H] ($\gtrsim$ $-$1.0) regime, and the other (dominant) group of NSCs ($\sim$ 63\%) run from the lower-[M/H] and older-age corner to the higher-[M/H] and younger-age end (i.e. lower right quadrant), and particularly they exhibit a broad negative age-[M/H] correlation. About $\sim$ 12\% of the NSCs with [M/H]$_{\rm diff}$ > 0 fall in the low-[M/H] and young-age quadrant of the diagram that cannot be assigned to the above two major groups.
It is remarkable that all of the NSCs in this minority group have relative low stellar mass.

To verify the negative age$-$[M/H] correlation mentioned above, we perform a ``Gaussian Mixture Model (GMM)'' decomposition of the age$-$[M/H] distribution of NSCs with [M/H]$_{\rm diff}$ > 0, by utilizing the python package \textsc{sklearn}. We find that most of the NSCs that constitute the visually identified dominant group belong to the same GMM cluster (data points enclosed by a green solid polygon in the right panels of Figure \ref{fig:agemetal-coloraam}) that is distinct from the rest of the NSCs. We perform a linear least-squares fitting to this dominant GMM cluster of NSCs and overplot the best-fit relation (green dash-dotted line). The corresponding Spearman rank correlation coefficient is $-$0.56, with a p-value of 0.0037, suggesting a significant negative correlation.


The age-[M/H] distributions are independent of galaxy stellar mass and NSC mass, except that most low mass galaxies (i.e. $\logmhost$ $\lesssim$ 8.5) fall into the subsample with [M/H]$_{\rm diff}$ < 0. The above-mentioned dominant group of NSCs with [M/H]$_{\rm diff}$ > 0 tend to have younger host ages and lower [$\alpha$/Fe] towards the higher-[M/H] end, albeit with substantial scatter, while the subdominant group of NSCs in the old-age, high-[M/H] quadrant mostly have relatively old host age and similarly low [$\alpha$/Fe] to the dominant group.

The negative age-metallicity relation for the majority of NSCs with [M/H]$_{\rm diff}$ > 0 is a reflection of extended period of self-enrichment of the NSCs in their host galaxies. Given the lack of synchrony between metal enrichment and mass growth already inferred in Section \ref{sec:metaldiff}, it is not surprising that there is no systematic difference of $M_{\rm NSC}$ along the age-metallicity relation. In contrast, the lack of an age-metallicity relation for NSCs with [M/H]$_{\rm diff}$ < 0 may be attributed to an early assembly of NSCs over relatively short timescales. With that said, the small fraction of [M/H]$_{\rm diff}$ < 0 NSCs with very young ages (i.e. $\lesssim$ 2 Gyr) may imply substantial in-situ star formation in rare circumstances \citep{Paudel2020}.

\begin{figure}
  \includegraphics[width=0.5\textwidth]{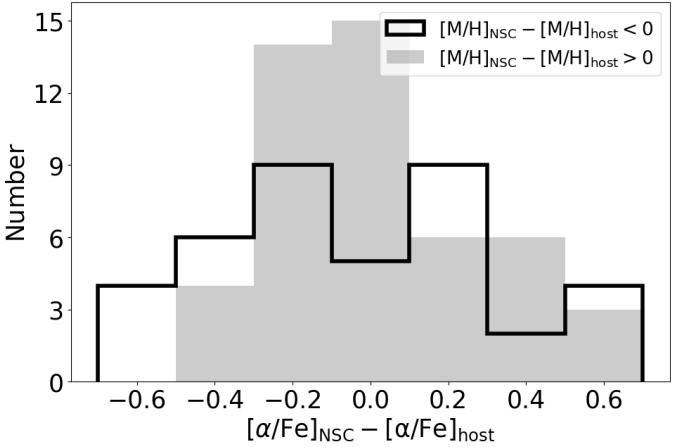}
  \caption{Distributions of [$\alpha$/Fe] difference between NSCs and the circum-NSC host. NSCs with [M/H]$_{\rm diff}$ > 0 and < 0 are shown separately, as indicated in the legend.}
  \label{fig:alpha2fe_diff}
\end{figure}

When gas inflow triggers in-situ star formation in NSCs, the same inflow event should have triggered nearly synchronous star formation in the circum-NSC host regions. Therefore, we expect that NSCs that grow more or less in synchrony with the circum-NSC host should have similar [$\alpha$/Fe] (an indicator of metal-enrichment timescales; see Section \ref{sec:alpha2fe_estimate}). The [$\alpha$/Fe]$_{\rm NSC}$ $-$ [$\alpha$/Fe]$_{\rm host}$ distributions of our sample are shown in Figure \ref{fig:alpha2fe_diff}. As expected, the distriution of NSCs with [M/H]$_{\rm diff}$ > 0 is highly peaked near [$\alpha$/Fe]$_{\rm NSC}$ $-$ [$\alpha$/Fe]$_{\rm host}$ $\sim$ 0.0, whereas NSCs with [M/H]$_{\rm diff}$ < 0 have much flatter distribution of [$\alpha$/Fe]$_{\rm NSC}$ $-$ [$\alpha$/Fe]$_{\rm host}$.

\subsection{Dependence on Hubble type of the host galaxies} \label{sec:dep_hubble}

\begin{figure}
  \includegraphics[width=0.5\textwidth]{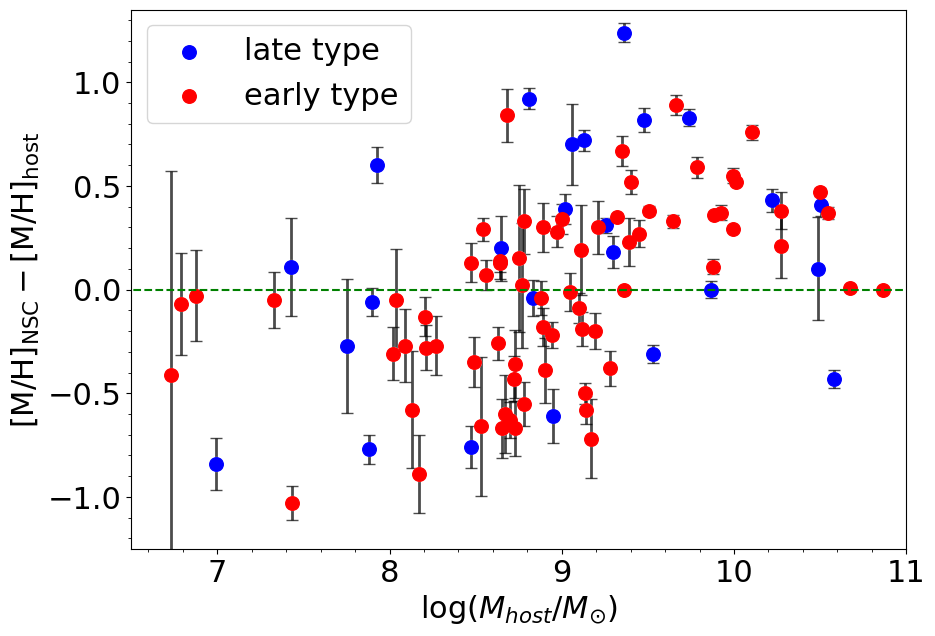}
  \caption{Similar to Figure \ref{fig:metaldiff}, except that the data points are differentiated according to the Hubble types of the galaxies, with early types being plotted in red and late types in blue.}
  \label{fig:metaldiff_hubble}
\end{figure}

\begin{figure*}
  \includegraphics[width=\textwidth]{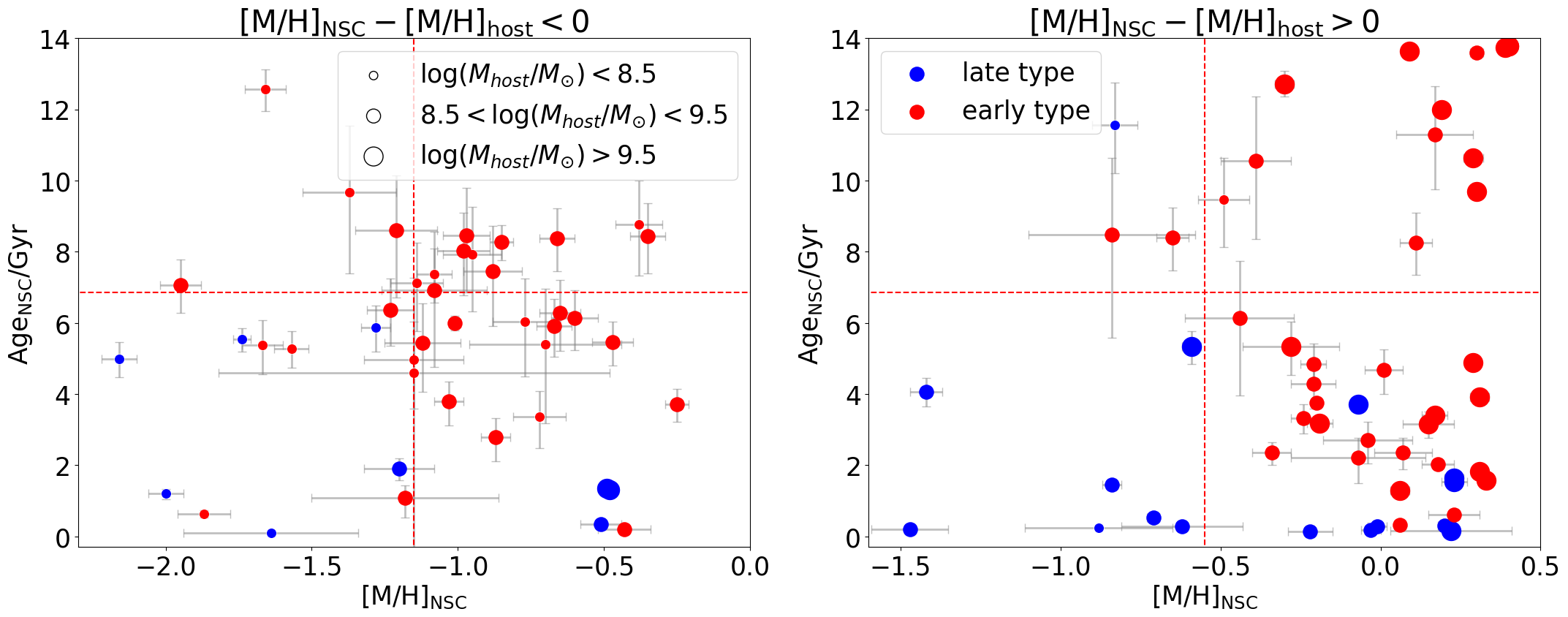}
  \caption{Similar to Figure \ref{fig:agemetal-coloraam}, except that the data points are differentiated according to the Hubble types of the galaxies, with early types being plotted in red and late types in blue.}
  \label{fig:metalage_hubble}
\end{figure*}

Here we explore if the stellar population properties of NSCs are related to the morphological type of the host galaxies. To this end, the [M/H]$_{\rm diff}$$-$$\logmhost$ distribution (Figure \ref{fig:metaldiff_hubble}) and the age$_{\rm NSC}$$-$[M/H]$_{\rm NSC}$ distribution (Figure \ref{fig:metalage_hubble}) are plotted by differentiating the host galaxies into early Hubble types (numerical Hubble type < 0; S0/a or earlier) and late Hubble types (numerical Hubble type > 0). Given the inhomogeneous nature of our sample, we do not attempt to perform a quantitative comparison of the distributions of different Hubble types in the explored parameter space, instead, we focus on the overall coverage of the parameter space and examine if there is systematic difference of NSC stellar populations between early and late types. 

The sample is dominated by early type hosts. However, the late types appear to cover similar parameter space to the early types in the [M/H]$_{\rm diff}$ vs. $\logmhost$ diagram. In the the age$_{\rm NSC}$ vs. [M/H]$_{\rm NSC}$ diagrams, it is remarkable that all of the NSCs with [M/H]$_{\rm diff}$ > 0 and occupy the lower-[M/H] and younger-age (lower left) quadrant reside in late type hosts with $\logmhost$ < 9.5, while all of the NSCs with [M/H]$_{\rm diff}$ > 0 and occupy the higher-[M/H] and older-age (upper right) quadrant reside in early type hosts. The rest of NSCs with [M/H]$_{\rm diff}$ > 0 in both early type and late type host appear to follow the negative age-[M/H] relation described in Section \ref{sec:agemetal}.

\subsection{Dependence on host galaxy environment}
\label{sec:galaxyenv}

\begin{figure}
    \centering
    \includegraphics[width=0.5\textwidth]{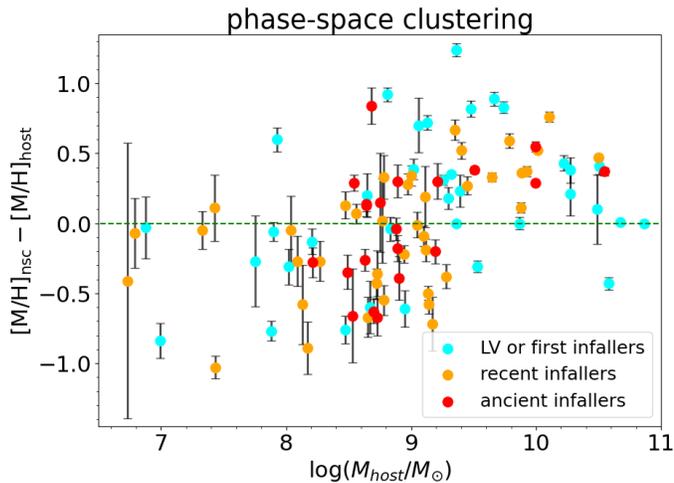}
    \caption{Similar to Figure \ref{fig:metaldiff}, except that the sample is divided into subsamples according to the phase-space classification (see Figure \ref{fig:phase2} and Section \ref{sec:ppsdiagram}).}
    \label{fig:metaldiff-env}
\end{figure}

\begin{figure*}
    \centering
    \includegraphics[width=\textwidth]{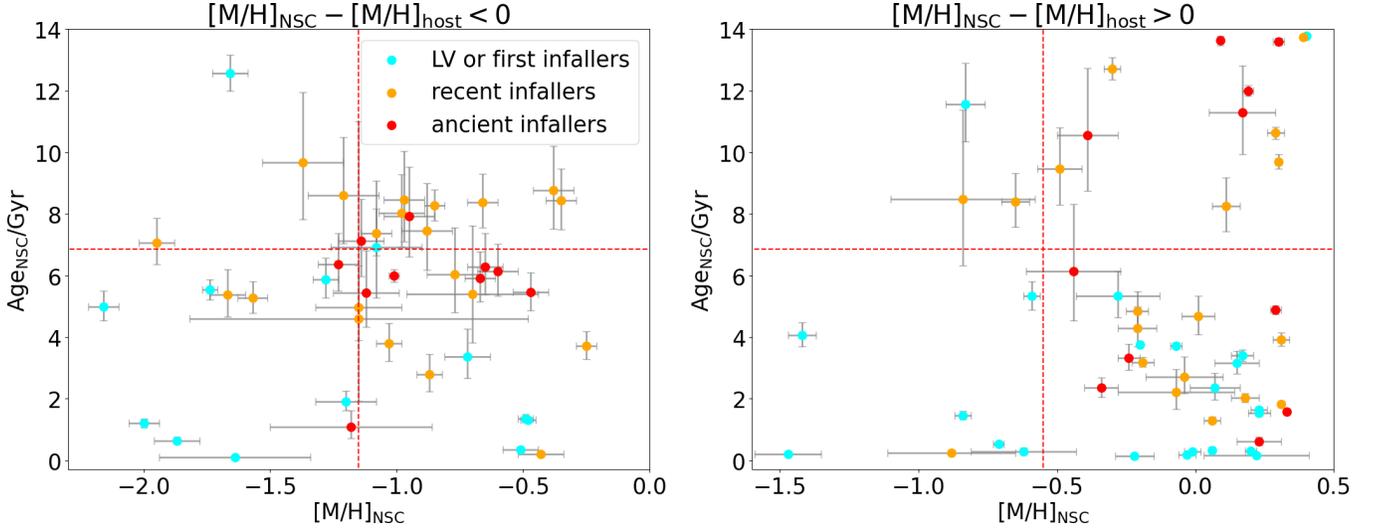}
    \caption{Similar to Figure \ref{fig:agemetal-coloraam}, except that the sample is divided into subsamples according to the phase-space classification (see Figure \ref{fig:phase2} and Section \ref{sec:ppsdiagram}), as indicated in the legend.}
    \label{fig:agemetal-pps}
\end{figure*}

Here we explore the environmental dependence of the stellar population properties of NSCs. According to Section \ref{sec:ppsdiagram}, our sample can be divided into subsamples based on the location in the projected phase-space. Each galaxy in the sample is assigned as one of the four environmental types of `ancient infallers', `recent infallers', `first infallers' and `LV' based on its location in the projected phase space diagram. 

We present the $ [\rm M/H]_{\rm diff} -  \logmhost$ relation by color-coding our galaxies according to their phase-space classifications in Figure~\ref{fig:metaldiff-env}. Note that the LV and first infallers types are combined as `LV or first infallers'. 
We find that different phase-space subsamples generally cover similar $ [\rm M/H]_{\rm diff} $ ranges for given galaxy stellar mass in Figure~\ref{fig:metaldiff-env}, except that there is a lack of ancient infallers at $\logmhost \lesssim $ 8.0 in our sample. We also note that NSCs with the highest $ [\rm M/H]_{\rm diff} $ values are mostly LV or first infallers. This apparent preference is probably due to the fact that the `LV or first infallers' sample is biased to late-type galaxies (Figure \ref{fig:metaldiff_hubble}) with relatively young NSC ages (Figure \ref{fig:metaldiff}). Given the heterogeneous nature of our sample, we do not attempt to further quantify the difference or similarity of different subsamples, but point out that no clear dependence on phase-space classifications is found in the $ [\rm M/H]_{\rm diff} -  \logmhost$ diagram.
%



\section{Discussion}
\label{sec:discussion}

\subsection{The formation mechanisms of NSCs revealed by stellar population studies}
\label{sec:5.1}
The plausible formation mechanisms of NSCs proposed in the literature broadly fall into two categories: 1) dynamical friction-driven inspiral and merger of either classical GCs or young star clusters and 2) in-situ star formation \citep[e.g.][]{Neumayer2020}. It is generally thought that in-situ star formation may be the dominant mechanism for relatively high-mass galaxies ($ \logmhost \gtrsim 9 $), while infall and mergers of star clusters may be more important for low mass galaxies ($ \logmhost<9 $). The theoretical models by \cite{Antonini2015} took in-situ star formation, GC infall process, as well as tidal influence of central massive black hole to NSCs into consideration and presented a co-evolution picture of NSCs, black holes and host galaxies. \cite{Antonini2015} found that for most host galaxies, the mass fraction of NSCs formed by in-situ star formation can be $ \sim 40\% $. The models reproduced the dropping nucleation fraction in high galaxy mass end but failed to explain the observed decrease of nucleation fraction toward the low galaxy mass end \citep{Sánchez-Janssen2019}. \cite{Paudel2011} analyzed spectra of 26 nucleated dE galaxies in the Virgo cluster and found most of their NSCs are significantly younger than galactic main bodies, with an average age difference of 3.5 Gyr, indicating gas accretion into NSCs. \cite{Paudel2011} also found fairly old and metal poor NSCs in very faint dEs, resembling the properties of their GC population. This suggests that NSCs in faint dEs might have formed by different processes than the NSCs in brighter dEs. \cite{Fahrion2021} and \cite{Fahrion2022b} together studied star formation histories of 34 nucleated galaxies from the Fornax cluster and Local Volume using MUSE IFU spectra. They quantified the mass fraction of NSCs likely formed through in-situ star formation by $ f_{\rm in-situ} $ and found a positive correlation with $ \logmhost $ and concluded that the transition of dominant NSC formation channels roughly occurs at $ \logmhost \sim 9 $ \citep{Fahrion2022a}. \cite{Kacharov2018} also studied 6 galaxies with high resolution optical spectra and found very young ($ <1Gyr $) stellar populations in their NSCs, suggesting prolonged in-situ star formation. \cite{Neumayer2020} presented the relation between metallicities and mass of 35 NSCs and their host galaxies with spectroscopic metallicities available in the literature by then, and noticed an apparent transition at $\logmhost \sim9 $. The \cite{Neumayer2020} spectroscopic sample covers a galaxy mass range of  $ 10^{8}M_{\odot} $ to $ 10^{9.7}M_{\odot} $. 

The present work improves upon previous spectroscopic studies of NSCs and their host galaxies by increasing the sample size by more than a factor of three with consistent treatment of spectral extraction and stellar population modeling. Our sample covers a galaxy stellar mass range of $ \logmhost \sim 6.5 $ to $ \logmhost \sim 11 $. 

We find the NSC metallicities have a stronger correlation with their host galaxy mass than NSC mass (Section \ref{sec:MZR}; see also  \cite{Fahrion2022b}). This suggests that NSCs coevolve with their host galaxies.  
At $ \logmhost>8.5 $, NSCs have increasingly higher average metallicities than typical red GCs in their host, implying that the GC infall-merger channel becomes more and more insignificant in more massive galaxies, while at $ \logmhost<8.5 $, NSCs have average metallicities close to that of typical red GCs but systematically below the circum-NSC host. This suggests that NSCs in low mass galaxies mainly formed through inspiral-merger of red GCs that are formed prior to the formation of the bulk of their circum-NSC host stellar populations. Alternatively, NSCs and red GCs may simply be two families of star clusters formed in a similar epoch but different spatial locations \citep{Sánchez-Janssen2019}.  In Section \ref{sec:model}, we will demonstrate the efficacy of GC inspiral-merger in producing NSCs in low mass galaxies.

The broad galaxy mass coverage of our sample enables a discovery of a broad peak of relative metal-enrichment of NSCs with respect to the cirum-NSC host regions near $\logmhost \sim 9.8$, beyond which the average NSC-host metallicity difference $ [\rm M/H]_{\rm diff} $ decreases with galaxy mass, reaching zero average difference near $\logmhost \sim$ 11. This indicates that in-situ star formation of NSCs is presently most active in these galaxies. 

The relatively large sample size also enables a discovery that a major group of NSCs with $ [\rm M/H]_{\rm diff} >0 $ follow a negative age$-$metallicity correlation (Figure \ref{fig:agemetal-coloraam}). This is a vivid evidence for extended period of metal enrichment of NSCs through in-situ star formation. An early intensive formation followed by a quick quenching may explain the subdominant group of NSCs with old age and high metallicities in the $ [\rm M/H]_{\rm diff} >0 $ subsample. No age$-$metallicity correlation is found for the subsample with $ [\rm M/H]_{\rm diff} <0 $.

\subsection{The contribution of dynamical friction-driven GC inspiral to NSC formation}
\label{sec:model}
The finding that the NSC metallicity becomes systematically lower than (similar to) circum-NSC host (typical red GCs) toward the lower galaxy mass end ($\logmhost \lesssim 9.5$; Figures \ref{fig:massmetal}, \ref{fig:metaldiff}) is in line with the GC inspiral-merger scenario of NSC formation. Dynamical friction is the key process that drives the inspiral of GCs.  In this section, we perform a test of the relevance of dynamical friction to the formation of NSCs with simplified models. By assuming a standard isothermal dark matter halo, the formula of dynamical friction time of a GC (from an initial radius $r_{i}$ to galaxy center) is reduced to \citep[]{Binney1987}
\begin{equation}\label{equ:tdf}
\footnotesize
t_{\rm DF} = \frac{2.64\times10^{2}}{\ln{\Lambda}}\left(\frac{r_{i}}{2~\rm{kpc}}\right)^{2}\left(\frac{v_{c}}{250~ \rm{km/s}}\right)\left(\frac{10^{6}~\rm{M_{\odot}}}{M}\right) \rm Gyr,
\end{equation}
where $v_{c}$ is the circular velocity of a GC with mass $M$ at radius $r_{i}$ of an isothermal galaxy halo. The Coulomb logarithm $\ln{\Lambda}$ is set to 10. With this formula, we determine whether or not a GC of given mass and initial galactocentric radius can spiral into galaxy centers in 10 Gyr. To set up the initial spatial distribution of GCs in galaxies of different mass, we adopt various observational scaling relations of galaxies available in the recent literature (see below), which allows us to evaluate the total number and mass of GCs that would spiral to galaxy center and contribute to the NSC assembly. The predicted NSC$-$galaxy mass relation can be compared to observations.

To construct the empirical model, we assume the host galaxies obey the observed effective radius size$-$mass relation \citep{Mowla2019}, the S\'ersic index$-$galaxy stellar mass relation \citep[][Equation 12]{Graham2006}, and the galaxy halo$-$stellar mass relation \citep[][Equation 66]{R.Puebla2017}. The total mass of the GC system of a galaxy is estimated based on its linear correlation with galaxy halo mass \citep[]{Harris2017}. In addition, the GC system is assumed to follow a Gaussian mass function, with a mean of 10$^{5.3}$ M$_{\odot}$ and logarithmic standard deviation of 0.5 \citep[]{Jordan2007}. To estimate the number of red GCs and blue GCs, we adopt the positive relation between the fraction of red GCs and galaxy stellar mass found by \cite{Peng2006}. To set up the spatial distribution of blue and red GCs, we adopt the recently observed blue and red GC system effective radius$-$galaxy halo mass relation \cite[][Equations 17, 18]{Lim2024}, and further assume the GC systems follow S\'ersic radial profiles with the same S\'ersic indices as galaxy star light. 


\begin{figure*}
  \centering
  \includegraphics[width=\textwidth]{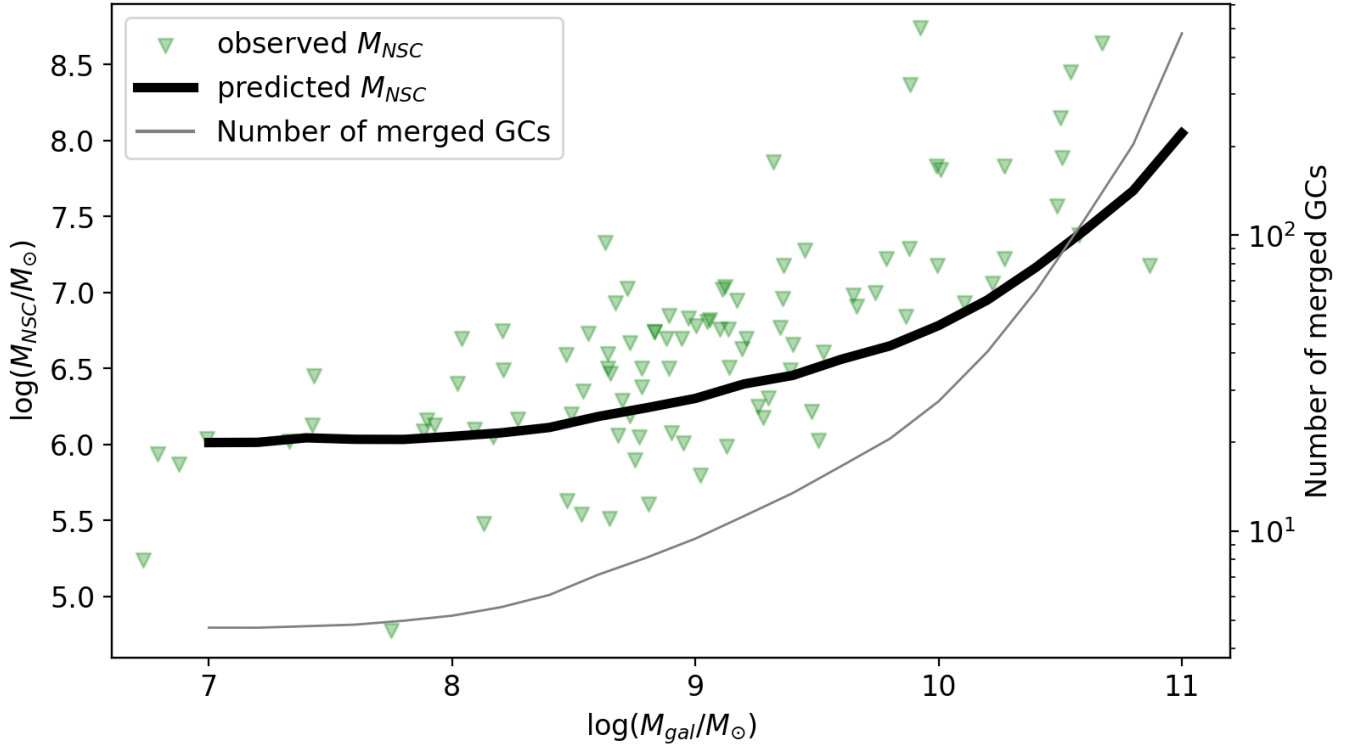}
  \caption{Comparison of the observed and predicted NSC$-$galaxy stellar mass relation. The green inverted triangles represent the observed sample studied in this work. The black (grey) curve is the average NSC$-$galaxy mass (number of merged GCs vs. galaxy mass) relation predicted by a model of NSC formation through dynamical friction-driven inspiral and merger. See Section \ref{sec:model} for details.}
  \label{fig:gcmodel}
\end{figure*}

With the above model setting, we estimate the number and mass of GCs that would spiral into the host center (and thus contribute to the predicted NSC mass) in 10 Gyr for a series of input galaxy stellar masses ranging from 10$^{7}$ to 10$^{11}$ M$_{\odot}$. A Monte Carlo method is adopted to randomly draw GCs (for 1000 times) according to the Gaussian GC mass function for given total GC mass and galactocentric radius. The predicted NSC$-$ and GC number$-$galaxy stellar mass relations are shown in Figure~\ref{fig:gcmodel}. The predicted average $ M_{\rm NSC} $ increases steadily with galaxy stellar mass, and slope of the relation gradually steepens at higher galaxy mass. In addition, the average number of merged GCs also increases with galaxy mass. Note that the estimation based on existing GCs may be literally regarded as the future growth potential of NSCs through dynamical friction. However, considering that dynamical friction-driven migration has been a continuous process, the growth potential also reflects the past growth efficiency driven by inspiral of {\it classical} GCs.

The predicted NSC$-$galaxy stellar mass relation is in remarkable agreement with the average trend of observational sample at $\logmhost$ $<$ 9.5, which is in line with the prevalent notion in the literature. At higher galaxy mass, the observed NSC masses are systematically higher than predictions. The agreement at low galaxy masses reinforces the notion that GC inspiral-merger is the dominant NSC formation mechanism in these galaxies.  

The relatively low predicted $ M_{\rm NSC} $ at $ \log M_{\rm gal} > 9.5 $ may suggest that NSC formation mechanisms other than GC inspiral play important roles in these galaxies. {\bf The systematic discrepancy may be also partly attributed to the ignorance of galaxy growth history in our simple model. The present-day high-mass galaxies may have experienced more significant and active growth (via in-situ star formation, accretion, or galaxy mergers) than lower mass galaxies and the overall inspiral efficiency of GCs in the high-mass galaxies could be higher in earlier epoch than that predicted by our model.} Nevertheless, it is noteworthy that $\logmhost$ $\sim$ 9.5 is also where we find systematically higher  $[\rm M/H]_{\rm NSC}$ than $[\rm M/H]_{\rm host}$ (e.g., Figure~\ref{fig:metaldiff}) toward higher galaxy mass, which implies an important role of in-situ star formation in the NSC formation of these relatively high mass galaxies. 

The NSC formation scenario of dynamical friction-driven cluster inspiral has been explored with theoretical models in the literature. However, most previous studies focused on exploring the possible parameter space $-$ such as the star cluster mass function and spatial distribution, galaxy structural properties $-$ that can explain
the observed NSCs within the framework of dynamical friction  scenario \citep[e.g.][]{Bekki2010,Arca-Sedda2014,Leaman2022}. Our analysis differs from previous studies by performing an explicit test of dynamical friction of classical GCs, based exclusively on
observed scaling relations established in the recent literature for GCs and galaxy structures, with virtually no free parameter. Such a simple and clean test is worthwhile, especially given the theoretical uncertainties such as star cluster formation, disruption efficiencies and the contribution of in-situ star formation to NSCs.

According to previous studies \citep[e.g.][]{Arca-Sedda2014}, cluster disruption effect is expected to be significant only in relatively massive galaxies. Therefore, the deviation of the predicted NSC mass from the observed ones in high mass galaxies would be even more significant if including the cluster disruption effect. In addition, our analysis ignores the plausible NSC erosion effect caused by massive black hole binaries located near the center of relatively massive galaxies \citep[e.g.][]{Antonini2015}, which would further reduce the NSC mass predicted by our models toward the high mass end.
Lastly, we point out that any star clusters (being young or old, gas-rich or gas-poor) that are massive
enough and survive long enough may be driven to galaxy center through dynamical friction. Therefore, our empirical model is far from being flawless, and there is no doubt that more
sophisticated models involving both star cluster formation and disruption histories within a hierarchical and continuous galaxy assembly framework is the way toward a complete understanding of NSC
formation.

\section{Conclusions}
\label{sec:conclusion}
To shed light on the formation mechanisms of nuclear star clusters (NSCs), we collect VLT optical spectroscopic observations of nuclear regions of 97 galaxies in the Local Group, Virgo cluster and Fornax cluster. Among the sample, 29 are first analyzed in this work. We perform uniform data processing, subtract underlying host galaxy spectra from NSCs, and derive the stellar population properties (mean age, mean metallicity, $ [\rm \alpha /Fe] $) of both NSCs and the circum-NSC  host regions. We explore the NSCs formation and growth mechanism by analyzing the mass-metallicity distribution, NSC$-$host metallicity difference $ [\rm M/H]_{\rm diff} $, age-metallicity relations, and the environmental dependence of the stellar population of NSCs. Our main results are summarized as follows:

(i) The NSC metallicities have a more significant correlation with their host galaxy mass than the NSC mass, suggesting that NSCs coevolve with their host galaxies.  At $ \logmhost \gtrsim 8.5 $, NSCs have increasingly higher average metallicities than typical red GCs towards the higher galaxy mass end, indicating that NSC formation mechanisms other than inspiral-merger of classical GCs become increasingly important toward higher galaxy mass. At $ \logmhost \lesssim 8.5 $, NSCs have average metallicities similar to or slightly lower (toward the lower galaxy mass end) than that of red GCs, implying GC inspiral-merger being the dominant formation channel of NSCs. Typical NSCs in the lowest mass galaxies may have been formed even prior to the formation of most GCs.

(ii) We identify three galaxy mass regimes in the $ [\rm M/H]_{\rm diff} $ $-$ galaxy stellar mass diagram. The three regimes are separated by $\logmhost$ of 8.5 and 9.5 respectively. In the low galaxy mass regime (i.e. $ \logmhost \lesssim 8.5 $), nearly all NSCs have lower metallicities than the circum-NSC host regions, while in the high mass regime ($ \logmhost \gtrsim 9.5 $), nearly all NSCs have higher metallicities than the circum-NSC host regions. In intermediate mass regime (8.5 $\lesssim$ $\logmhost$ $\lesssim$ 9.5), the fraction of NSCs with higher metallicities than the circum-NSC host is comparable to those with lower metallicities than circum-NSC host. These findings suggest that NSC growth is sustained by continuous metal-enriched gas inflow from larger galactocentric distances in high-mass galaxies, while in low-mass galaxies enriched gas inflow to the NSC region has been largely suppressed with respect to the circum-host regions. In the intermediate mass regime, larger $ [\rm M/H]_{\rm diff} $ is primarily due to higher NSC metallicities rather than lower circum-host metallicities. The growth of NSCs through enriched gas inflow in intermediate mass galaxies is probably sporadic in time, which results in a lack of correlation between $ [\rm M/H]_{\rm diff} $ and NSC mass.

(iii) In the intermediate and high galaxy mass regimes, the average $ [\rm M/H]_{\rm diff} $ reaches a broad maximum at $\logmhost$ $\sim$ 9.8, and drops toward both the higher and lower galaxy mass end. Galaxies with the highest $ [\rm M/H]_{\rm diff} $ (i.e. $>$ 0.5) are mostly characterized by relatively low circum-NSC metallicities (instead of systematically high NSC metallicities), young NSC ages and low NSC mass, suggesting that NSCs in these galaxies are at an relatively early stage of assembly through in-situ star formation. Growth of NSCs towards higher galaxy mass end is probably subjected to significant negative influence either through tidal disruption of supermassive black holes or quenching by active galactic nuclei.

(iv) By dividing NSCs into subsamples with $ [\rm M/H]_{\rm diff} $ $<$ 0 or $>$ 0, we find that the $ [\rm M/H]_{\rm diff} $ $<$ 0 subsample of NSCs have virtually no relation between light-weighted age and metallicity, while for the majority of NSCs in the $ [\rm M/H]_{\rm diff} $ $>$ 0 subsample there exists a negative age$-$metallicity correlation, irrespective of galaxy stellar mass, in line with an extended chemical enrichment history. A synchrony between the metal-enrichment history of NSCs and circum-NSC host for the $ [\rm M/H]_{\rm diff} $ $>$ 0 subsample is also reflected in a relatively high fraction of NSCs with similar [$\alpha$/Fe] to the circum-NSC host. 

(v) A simplified empirical model of NSC formation through dynamical friction-driven inspiral-merger of GCs based exclusively on various observed scaling relations of present-day galaxies explains the average NSC$-$galaxy mass relation at $\logmhost < 9.5$, in agreement with similar studies in the literature, implying that other formation mechanisms are inconsequential to NSC formation in dwarf  galaxies. We find that, about 10 or so GCs are sufficient to double the NSC mass in these low mass galaxies. The empirical model fails to explain $ M_{\rm NSC} $ at $ \log M_{\rm gal} > 9.5 $, which coincides with the ``transition'' mass where NSC metallicities are systematically higher than the circum-NSC host, reinforcing that in-situ star formation plays an important role in the assembly of NSCs of high mass galaxies.

With the limited sample size in mind, we find no significant environmental dependence of the stellar population properties of NSCs, and no significant dependence on the Hubble type of host galaxies. Our homogeneous analysis of the stellar population properties of nearby NSCs reinforce the notion of a strong galaxy mass dependence of the assembly of NSCs, with an seemingly abrupt transition near $\logmhost \sim 9.0\pm 0.5$.

\section*{Acknowledgements}

We thank the anonymous referee for helpful suggestions that improved our manuscript. We acknowledge support from the NSFC grant (Nos. 12122303, 11973039, 11421303, 11973038, 12233008). This work is also supported by the National Key Research and Development Program of China (2023YFA1608100) and the China Manned Space Project (Nos.CMS-CSST-2021-B02, CMS-CSST-2021-A07). We acknowledge support from the CAS Pioneer Hundred Talents Program, the Strategic Priority Research Program of Chinese Academy of Sciences (Grant No. XDB 41000000) and the Cyrus Chun Ying Tang Foundations.
We also acknowledge support from the Mid-career Researcher Program (No. RS-2023-00208957).

 
\bibliography{reference}

\appendix
\section{S/N of spectrum}
The spectral S/N of NSCs and circum-host of our sample galaxies are given in Table~\ref{tab:sn}.

\begin{table}
    \caption{S/N of spectra at ~5000 \AA. ${S/N}_{\rm NSC}$ and ${S/N}_{\rm host}$ are the signal to noise ratio of NSCs spectra and host galaxies spectra before being smoothed to FORS2 resolution while the col 'instrument' explain the observation instrument of the spectra.}
    \label{tab:sn}
    \hspace{-1cm}
    \begin{tabular}{cccc}
    \hline
    Gal & ${S/N}_{\rm NSC}$ & ${S/N}_{\rm host}$ & instrument \\
    \hline
    NGC247 & 112.76 & 24.57 & X-shooter \\
    NGC2784 & 74.43 & 123.83 & X-shooter \\
    NGC300 & 19.55 & 21.07 & X-shooter \\
    NGC3115 & 171.63 & 201.35 & X-shooter \\
    NGC3621 & 42.29 & 132.6 & X-shooter \\
    NGC5068 & 53.38 & 38.59 & X-shooter \\
    NGC5102 & 133.5 & 114.29 & X-shooter \\
    NGC5206 & 79.46 & 68.79 & X-shooter \\
    NGC5236 & 44.81 & 60.72 & X-shooter \\
    NGC628 & 53.15 & 40.6 & X-shooter \\
    NGC7713 & 26.45 & 17.83 & X-shooter \\
    NGC7793 & 108.89 & 26.61 & X-shooter \\
    VCC0216 & 11.91 & 14.58 & FORS2 \\
    VCC0308 & 7.09 & 18.7 & FORS2 \\
    VCC0389 & 8.12 & 13.15 & FORS2 \\
    VCC0490 & 10.49 & 13.53 & FORS2 \\
    VCC0545 & 11.72 & 10.83 & FORS2 \\
    VCC0592 & 6.09 & 12.72 & FORS2 \\
    VCC0725 & 9.17 & 6.45 & FORS2 \\
    VCC0765 & 8.05 & 15.98 & FORS2 \\
    VCC0786 & 2.77 & 5.61 & MUSE \\
    VCC0856 & 9.69 & 11.65 & FORS2 \\
    VCC0871 & 3.9 & 11.29 & FORS2 \\
    VCC0916 & 7.47 & 8.85 & FORS2 \\
    VCC0929 & 9.4 & 12.91 & FORS2 \\
    VCC0940 & 8.57 & 10.48 & FORS2 \\
    VCC0965 & 10.93 & 11.24 & FORS2 \\
    VCC0990 & 5.68 & 16.13 & FORS2 \\
    VCC1069 & 5.11 & 10.42 & FORS2 \\
    VCC1073 & 1.9 & 12.45 & FORS2 \\
    VCC1104 & 3.96 & 8.67 & FORS2 \\
    VCC1122 & 5.46 & 8.17 & FORS2 \\
    VCC1167 & 16.19 & 10.0 & FORS2 \\
    VCC1185 & 7.3 & 9.61 & FORS2 \\
    VCC1254 & 12.96 & 8.25 & FORS2 \\
    VCC1261 & 11.18 & 17.35 & FORS2 \\
    VCC1304 & 4.94 & 9.5 & FORS2 \\
    VCC1308 & 6.59 & 15.36 & FORS2 \\
    VCC1333 & 16.46 & 3.95 & FORS2 \\
    
    \hline
    \end{tabular}
\end{table}

\begin{table}
    \begin{tabular}{cccc}
    \hline
    Gal & ${S/N}_{\rm NSC}$ & ${S/N}_{\rm host}$ & instrument \\
    \hline
    VCC1348 & 19.65 & 9.64 & FORS2 \\
    VCC1353 & 10.43 & 12.48 & FORS2 \\
    VCC1355 & 8.6 & 11.37 & FORS2 \\
    VCC1386 & 2.89 & 14.54 & FORS2 \\
    VCC1389 & 11.25 & 10.46 & FORS2 \\
    VCC1407 & 11.31 & 17.94 & FORS2 \\
    VCC1431 & 9.08 & 12.47 & FORS2 \\
    VCC1491 & 2.8 & 1.99 & FORS2 \\
    VCC1661 & 14.07 & 7.18 & FORS2 \\
    VCC1826 & 6.61 & 15.19 & FORS2 \\
    VCC1861 & 10.02 & 11.79 & FORS2 \\
    VCC1945 & 7.98 & 9.73 & FORS2 \\
    VCC2019 & 8.21 & 14.22 & FORS2 \\
    CIRCINUS & 0.38 & 0.21 & MUSE \\
    ESO59-01 & 26.64 & 12.28 & MUSE \\
    FCC119 & 10.07 & 18.9 & MUSE \\
    FCC148 & 15.11 & 17.38 & MUSE \\
    FCC153 & 12.66 & 14.87 & MUSE \\
    FCC170 & 22.26 & 22.71 & MUSE \\
    FCC177 & 13.79 & 15.48 & MUSE \\
    FCC182 & 11.82 & 17.35 & MUSE \\
    FCC188 & 18.49 & 6.9 & MUSE \\
    FCC190 & 14.04 & 19.42 & MUSE \\
    FCC193 & 14.8 & 23.92 & MUSE \\
    FCC202 & 15.56 & 19.26 & MUSE \\
    FCC207 & 11.93 & 9.36 & MUSE \\
    FCC211 & 12.56 & 13.45 & MUSE \\
    FCC215 & 13.66 & 3.24 & MUSE \\
    FCC222 & 11.06 & 16.29 & MUSE \\
    FCC223 & 13.91 & 7.2 & MUSE \\
    FCC227 & 0.19 & 1.6 & MUSE \\
    FCC245 & 0.11 & 2.3 & MUSE \\
    FCC249 & 19.14 & 25.89 & MUSE \\
    FCC255 & 11.86 & 18.49 & MUSE \\
    FCC277 & 14.85 & 25.57 & MUSE \\
    FCC301 & 13.17 & 21.41 & MUSE \\
    FCC306 & 7.05 & 19.48 & MUSE \\
    FCC310 & 11.2 & 20.13 & MUSE \\
    FCC47 & 22.54 & 23.57 & MUSE \\
    FCCB1241 & 1.45 & 3.18 & MUSE \\
    IC1959 & 35.00 & 30.21 & MUSE \\
    IC5332 & 11.78 & 17.98 & MUSE \\
    KK197 & 7.34 & 6.39 & MUSE \\
    KKS58 & 13.81 & 5.82 & MUSE \\
    \hline
    \end{tabular}
\end{table}

\begin{table}
    \begin{tabular}{cccc}
    \hline
    Gal & ${S/N}_{\rm NSC}$ & ${S/N}_{\rm host}$ & instrument \\
    \hline
    NGC1487 & 20.59 & 40.92 & MUSE \\
    NGC1705 & 17.14 & 2.64 & MUSE \\
    NGC1796 & 3.97 & 11.75 & MUSE \\
    NGC2835 & 17.71 & 15.39 & MUSE \\
    NGC3274 & 2.93 & 0.62 & MUSE \\
    NGC3368 & 13.89 & 18.05 & MUSE \\
    NGC3489 & 16.17 & 15.19 & MUSE \\
    NGC3593 & 2.52 & 13.4 & MUSE \\
    NGC4592 & 11.04 & 25.22 & MUSE \\
    NGC5253 & 0.26 & 0.1 & MUSE \\
    NGC853 & 10.94 & 1.84 & MUSE \\
    UGC3755 & 0.21 & 0.15 & MUSE \\
    UGC5889 & 11.95 & 14.09 & MUSE \\
    UGC8041 & 21.24 & 12.13 & MUSE \\
    \hline
    \end{tabular}
\end{table}

\section{Light-weighted stellar population properties of NSCs and circum-NSC host of our sample galaxies}
Light-weighted age and light-weighted metallicity as well as $ [\alpha/Fe] $ of NSCs and circum-NSC host regions are given in Table~\ref{tab:ppxf result1}, Table~\ref{tab:ppxf result2} and Table~\ref{tab:ppxf result3} with their 1 $\sigma$ error.

\begin{table*}
        \caption{Stellar population properties of NSC and circum-NSC host regions.}
	\label{tab:ppxf result1}
        \hspace{-1.5cm}
        \renewcommand\arraystretch{1.5}
        \setlength{\tabcolsep}{3mm}{
	\begin{tabular}{ccccccc}
        \hline
        $\rm Gal$ & $ Age_{\rm NSC} $ (Gyr) & $[\rm M/\rm H]_{\rm NSC}$ & $[\alpha/\rm Fe]_{\rm NSC}$ & $Age_{\rm host}$ (Gyr) & $[\rm M/\rm H]_{\rm host}$ & $[\alpha/\rm Fe]_{\rm host}$
        \\
        \hline
        NGC247 & $0.53^{+0.03}_{-0.03}$ & $-0.71^{+0.02}_{-0.02}$ & $-0.16^{+0.08}_{-0.08}$ & $2.74^{+0.16}_{-0.15}$ & $-1.02^{+0.03}_{-0.03}$ & $1.13^{+0.14}_{-0.14}$ \\
        NGC2784 & $13.77^{+0.02}_{-0.02}$ & $0.4^{+0.0}_{-0.0}$ & $0.22^{+0.02}_{-0.02}$ & $11.06^{+0.1}_{-0.1}$ & $0.39^{+0.01}_{-0.01}$ & $0.36^{+0.01}_{-0.01}$ \\
        NGC300 & $1.46^{+0.13}_{-0.12}$ & $-0.84^{+0.03}_{-0.03}$ & $0.42^{+0.13}_{-0.13}$ & $5.92^{+0.55}_{-0.51}$ & $-1.56^{+0.04}_{-0.04}$ & $0.63^{+0.1}_{-0.1}$ \\
        NGC3115 & $13.78^{+0.02}_{-0.02}$ & $0.4^{+0.0}_{-0.0}$ & $0.45^{+0.01}_{-0.01}$ & $12.14^{+0.03}_{-0.03}$ & $0.4^{+0.0}_{-0.0}$ & $0.31^{+0.0}_{-0.0}$ \\
        NGC3621 & $1.64^{+0.09}_{-0.08}$ & $0.23^{+0.03}_{-0.03}$ & $-0.01^{+0.06}_{-0.06}$ & $1.7^{+0.08}_{-0.08}$ & $-0.6^{+0.03}_{-0.03}$ & $0.17^{+0.03}_{-0.03}$ \\
        NGC5068 & $0.29^{+0.02}_{-0.02}$ & $-0.01^{+0.03}_{-0.03}$ & $0.4^{+0.1}_{-0.1}$ & $1.15^{+0.15}_{-0.14}$ & $-0.83^{+0.05}_{-0.05}$ & $-0.12^{+0.06}_{-0.06}$ \\
        NGC5102 & $0.32^{+0.0}_{-0.0}$ & $0.06^{+0.01}_{-0.01}$ & $0.16^{+0.06}_{-0.06}$ & $0.92^{+0.04}_{-0.04}$ & $-0.29^{+0.01}_{-0.01}$ & $-0.04^{+0.01}_{-0.01}$ \\
        NGC5206 & $3.76^{+0.07}_{-0.07}$ & $-0.2^{+0.01}_{-0.01}$ & $0.02^{+0.02}_{-0.02}$ & $2.96^{+0.13}_{-0.13}$ & $-0.2^{+0.01}_{-0.01}$ & $-0.06^{+0.02}_{-0.02}$ \\
        NGC5236 & $1.32^{+0.07}_{-0.07}$ & $-0.48^{+0.03}_{-0.03}$ & $0.8^{+0.02}_{-0.02}$ & $0.73^{+0.03}_{-0.03}$ & $-0.05^{+0.03}_{-0.03}$ & $0.62^{+0.03}_{-0.03}$ \\
        NGC628 & $3.72^{+0.09}_{-0.09}$ & $-0.07^{+0.02}_{-0.02}$ & $0.26^{+0.08}_{-0.08}$ & $6.5^{+0.77}_{-0.69}$ & $-0.5^{+0.05}_{-0.05}$ & $-0.09^{+0.01}_{-0.01}$ \\
        NGC7713 & $0.19^{+0.01}_{-0.01}$ & $-0.03^{+0.03}_{-0.03}$ & $0.17^{+0.05}_{-0.05}$ & $2.83^{+0.21}_{-0.19}$ & $-0.95^{+0.04}_{-0.04}$ & $0.08^{+0.07}_{-0.07}$ \\
        NGC7793 & $0.3^{+0.01}_{-0.01}$ & $0.2^{+0.02}_{-0.02}$ & $0.07^{+0.03}_{-0.03}$ & $1.66^{+0.11}_{-0.11}$ & $-1.04^{+0.04}_{-0.04}$ & $0.07^{+0.04}_{-0.04}$ \\
        VCC0216 & $2.78^{+0.67}_{-0.54}$ & $-0.87^{+0.05}_{-0.05}$ & $-0.08^{+0.03}_{-0.03}$ & $4.18^{+0.72}_{-0.61}$ & $-0.49^{+0.07}_{-0.07}$ & $0.78^{+0.05}_{-0.05}$ \\
        VCC0308 & $2.36^{+0.48}_{-0.4}$ & $0.07^{+0.09}_{-0.09}$ & $0.29^{+0.05}_{-0.05}$ & $2.61^{+0.52}_{-0.43}$ & $-0.16^{+0.07}_{-0.07}$ & $-0.08^{+0.04}_{-0.04}$ \\
        VCC0389 & $2.22^{+0.73}_{-0.55}$ & $-0.07^{+0.21}_{-0.21}$ & $-0.09^{+0.02}_{-0.02}$ & $7.64^{+0.8}_{-0.73}$ & $-0.26^{+0.05}_{-0.05}$ & $-0.01^{+0.02}_{-0.02}$ \\
        VCC0490 & $8.43^{+1.04}_{-0.93}$ & $-0.35^{+0.06}_{-0.06}$ & $-0.28^{+0.01}_{-0.01}$ & $6.1^{+0.61}_{-0.55}$ & $-0.26^{+0.04}_{-0.04}$ & $0.23^{+0.04}_{-0.04}$ \\
        VCC0545 & $8.4^{+0.93}_{-0.83}$ & $-0.65^{+0.05}_{-0.05}$ & $0.49^{+0.05}_{-0.05}$ & $10.12^{+1.09}_{-0.98}$ & $-0.72^{+0.05}_{-0.05}$ & $0.16^{+0.08}_{-0.08}$ \\
        VCC0592 & $9.68^{+2.28}_{-1.85}$ & $-1.37^{+0.16}_{-0.16}$ & $0.15^{+0.01}_{-0.01}$ & $3.65^{+0.72}_{-0.6}$ & $-0.48^{+0.1}_{-0.1}$ & $0.38^{+0.0}_{-0.0}$ \\
        VCC0725 & $6.03^{+1.53}_{-1.22}$ & $-0.77^{+0.11}_{-0.11}$ & - & $9.52^{+1.59}_{-1.36}$ & $-0.72^{+0.08}_{-0.08}$ & - \\
        VCC0765 & $8.76^{+1.44}_{-1.24}$ & $-0.38^{+0.08}_{-0.08}$ & $0.75^{+0.02}_{-0.02}$ & $4.8^{+1.09}_{-0.89}$ & $-0.11^{+0.12}_{-0.12}$ & $-0.0^{+0.0}_{-0.0}$ \\
        VCC0786 & $8.6^{+1.89}_{-1.55}$ & $-1.21^{+0.14}_{-0.14}$ & - & $7.55^{+1.66}_{-1.36}$ & $-0.49^{+0.13}_{-0.13}$ & - \\
        VCC0856 & $4.67^{+0.66}_{-0.58}$ & $0.01^{+0.06}_{-0.06}$ & $-0.09^{+0.01}_{-0.01}$ & $10.13^{+0.87}_{-0.81}$ & $-0.66^{+0.04}_{-0.04}$ & $0.12^{+0.03}_{-0.03}$ \\
        VCC0871 & $1.08^{+0.54}_{-0.36}$ & $-1.18^{+0.32}_{-0.32}$ & $0.61^{+0.05}_{-0.05}$ & $6.52^{+1.55}_{-1.26}$ & $-0.52^{+0.1}_{-0.1}$ & $-0.12^{+0.0}_{-0.0}$ \\
        VCC0916 & $13.6^{+0.12}_{-0.12}$ & $0.3^{+0.02}_{-0.02}$ & $-0.03^{+0.0}_{-0.0}$ & $10.12^{+0.72}_{-0.67}$ & $0.16^{+0.05}_{-0.05}$ & $0.18^{+0.0}_{-0.0}$ \\
        VCC0929 & $8.26^{+0.92}_{-0.83}$ & $0.11^{+0.05}_{-0.05}$ & $0.22^{+0.02}_{-0.02}$ & $10.37^{+0.96}_{-0.88}$ & $-0.16^{+0.04}_{-0.04}$ & $0.74^{+0.01}_{-0.01}$ \\
        VCC0940 & $5.45^{+0.65}_{-0.58}$ & $-0.47^{+0.07}_{-0.07}$ & $-0.18^{+0.01}_{-0.01}$ & $7.56^{+0.62}_{-0.57}$ & $-0.43^{+0.04}_{-0.04}$ & $0.18^{+0.0}_{-0.0}$ \\
        VCC0965 & $2.35^{+0.34}_{-0.3}$ & $-0.34^{+0.06}_{-0.06}$ & $-0.37^{+0.02}_{-0.02}$ & $4.31^{+0.48}_{-0.43}$ & $-0.47^{+0.05}_{-0.05}$ & $-0.21^{+0.0}_{-0.0}$ \\
        VCC0990 & $4.29^{+0.7}_{-0.6}$ & $-0.21^{+0.07}_{-0.07}$ & $-0.08^{+0.03}_{-0.03}$ & $7.1^{+0.44}_{-0.41}$ & $-0.49^{+0.03}_{-0.03}$ & $-0.08^{+0.01}_{-0.01}$ \\
        
        \hline
	\end{tabular}}
\vspace{-3pt}
\end{table*}

\begin{table*}
	\label{tab:ppxf result2}
        \hspace{-1.5cm}
        \renewcommand\arraystretch{1.5}
        \setlength{\tabcolsep}{3mm}{
	\begin{tabular}{ccccccc}
        \hline
        $\rm Gal$ & $ Age_{\rm NSC} $ (Gyr) & $[\rm M/\rm H]_{\rm NSC}$ & $[\alpha/\rm Fe]_{\rm NSC}$ & $Age_{\rm host}$ (Gyr) & $[\rm M/\rm H]_{\rm host}$ & $[\alpha/\rm Fe]_{\rm host}$
        \\
        \hline
        VCC1069 & $4.96^{+1.37}_{-1.07}$ & $-1.15^{+0.17}_{-0.17}$ & $0.21^{+0.03}_{-0.03}$ & $8.74^{+0.85}_{-0.78}$ & $-0.88^{+0.05}_{-0.05}$ & $-0.23^{+0.0}_{-0.0}$ \\
        VCC1073 & $11.29^{+1.53}_{-1.35}$ & $0.17^{+0.12}_{-0.12}$ & $0.67^{+0.0}_{-0.0}$ & $12.71^{+0.7}_{-0.67}$ & $-0.13^{+0.04}_{-0.04}$ & $0.1^{+0.0}_{-0.0}$ \\
        VCC1104 & $7.93^{+1.6}_{-1.33}$ & $-0.95^{+0.1}_{-0.1}$ & $0.1^{+0.03}_{-0.03}$ & $5.34^{+0.66}_{-0.59}$ & $-0.6^{+0.07}_{-0.07}$ & $-0.11^{+0.0}_{-0.0}$ \\
        VCC1122 & $2.7^{+0.66}_{-0.53}$ & $-0.04^{+0.14}_{-0.14}$ & $-0.02^{+0.0}_{-0.0}$ & $10.64^{+1.4}_{-1.24}$ & $-0.37^{+0.07}_{-0.07}$ & $0.15^{+0.01}_{-0.01}$ \\
        VCC1167 & $8.02^{+1.25}_{-1.08}$ & $-0.98^{+0.09}_{-0.09}$ & $0.25^{+0.08}_{-0.08}$ & $10.44^{+1.14}_{-1.02}$ & $-0.55^{+0.06}_{-0.06}$ & $0.05^{+0.04}_{-0.04}$ \\
        VCC1185 & $5.44^{+1.38}_{-1.1}$ & $-1.12^{+0.13}_{-0.13}$ & $-0.3^{+0.14}_{-0.14}$ & $11.42^{+0.74}_{-0.7}$ & $-0.45^{+0.03}_{-0.03}$ & $-0.22^{+0.09}_{-0.09}$ \\
        VCC1254 & $3.71^{+0.48}_{-0.43}$ & $-0.25^{+0.04}_{-0.04}$ & $0.12^{+0.01}_{-0.01}$ & $6.7^{+0.82}_{-0.73}$ & $-0.06^{+0.07}_{-0.07}$ & $0.24^{+0.03}_{-0.03}$ \\
        VCC1261 & $2.02^{+0.15}_{-0.14}$ & $0.18^{+0.05}_{-0.05}$ & $-0.51^{+0.14}_{-0.14}$ & $4.11^{+0.42}_{-0.38}$ & $-0.34^{+0.03}_{-0.03}$ & $0.11^{+0.01}_{-0.01}$ \\
        VCC1304 & $6.92^{+2.15}_{-1.64}$ & $-1.08^{+0.18}_{-0.18}$ & $0.35^{+0.09}_{-0.09}$ & $5.53^{+0.92}_{-0.79}$ & $-0.48^{+0.05}_{-0.05}$ & $-0.27^{+0.01}_{-0.01}$ \\
        VCC1308 & $3.33^{+0.44}_{-0.39}$ & $-0.24^{+0.04}_{-0.04}$ & $0.41^{+0.04}_{-0.04}$ & $6.32^{+0.43}_{-0.4}$ & $-0.53^{+0.04}_{-0.04}$ & $-0.07^{+0.02}_{-0.02}$ \\
        VCC1333 & $7.37^{+0.8}_{-0.72}$ & $-1.08^{+0.06}_{-0.06}$ & $0.37^{+0.04}_{-0.04}$ & $7.03^{+3.65}_{-2.4}$ & $-1.03^{+0.24}_{-0.24}$ & $0.78^{+1.78}_{-1.78}$ \\
        VCC1348 & $5.91^{+0.87}_{-0.76}$ & $-0.67^{+0.06}_{-0.06}$ & $0.2^{+0.02}_{-0.02}$ & $11.74^{+1.15}_{-1.05}$ & $-0.41^{+0.05}_{-0.05}$ & $0.35^{+0.02}_{-0.02}$ \\
        VCC1353 & $3.37^{+0.89}_{-0.71}$ & $-0.72^{+0.09}_{-0.09}$ & $-0.08^{+0.03}_{-0.03}$ & $2.55^{+0.58}_{-0.47}$ & $-0.41^{+0.09}_{-0.09}$ & $0.49^{+0.04}_{-0.04}$ \\
        VCC1355 & $6.29^{+1.07}_{-0.91}$ & $-0.65^{+0.07}_{-0.07}$ & $-0.23^{+0.03}_{-0.03}$ & $5.35^{+1.8}_{-1.34}$ & $-0.26^{+0.14}_{-0.14}$ & $-0.01^{+0.05}_{-0.05}$ \\
        VCC1386 & $10.55^{+2.19}_{-1.81}$ & $-0.39^{+0.11}_{-0.11}$ & - & $10.05^{+1.01}_{-0.92}$ & $-0.69^{+0.04}_{-0.04}$ & - \\
        VCC1389 & $7.12^{+1.36}_{-1.14}$ & $-1.14^{+0.09}_{-0.09}$ & $0.13^{+0.11}_{-0.11}$ & $7.63^{+0.94}_{-0.84}$ & $-0.86^{+0.06}_{-0.06}$ & $-0.09^{+0.02}_{-0.02}$ \\
        VCC1407 & $6.37^{+1.01}_{-0.87}$ & $-1.23^{+0.08}_{-0.08}$ & $0.12^{+0.09}_{-0.09}$ & $9.91^{+0.57}_{-0.54}$ & $-0.6^{+0.03}_{-0.03}$ & $0.14^{+0.02}_{-0.02}$ \\
        VCC1431 & $6.13^{+0.9}_{-0.79}$ & $-0.6^{+0.08}_{-0.08}$ & $0.59^{+0.01}_{-0.01}$ & $11.05^{+0.98}_{-0.9}$ & $-0.4^{+0.03}_{-0.03}$ & $0.42^{+0.0}_{-0.0}$ \\
        VCC1491 & $6.14^{+2.17}_{-1.6}$ & $-0.44^{+0.17}_{-0.17}$ & $0.56^{+0.02}_{-0.02}$ & $6.03^{+2.94}_{-1.98}$ & $-0.59^{+0.31}_{-0.31}$ & $-0.43^{+0.03}_{-0.03}$ \\
        VCC1661 & $8.45^{+1.6}_{-1.34}$ & $-0.97^{+0.08}_{-0.08}$ & $-0.04^{+0.02}_{-0.02}$ & $6.59^{+1.65}_{-1.32}$ & $-0.3^{+0.12}_{-0.12}$ & $-0.07^{+0.02}_{-0.02}$ \\
        VCC1826 & $9.47^{+1.33}_{-1.17}$ & $-0.49^{+0.08}_{-0.08}$ & $-0.27^{+0.06}_{-0.06}$ & $9.03^{+0.82}_{-0.75}$ & $-0.62^{+0.05}_{-0.05}$ & $-0.19^{+0.02}_{-0.02}$ \\
        VCC1861 & $8.38^{+0.92}_{-0.83}$ & $-0.66^{+0.06}_{-0.06}$ & $-0.2^{+0.02}_{-0.02}$ & $7.25^{+0.53}_{-0.49}$ & $-0.08^{+0.03}_{-0.03}$ & $0.03^{+0.02}_{-0.02}$ \\
        VCC1945 & $7.45^{+1.54}_{-1.28}$ & $-0.88^{+0.1}_{-0.1}$ & $0.63^{+0.14}_{-0.14}$ & $6.41^{+1.41}_{-1.15}$ & $-0.52^{+0.13}_{-0.13}$ & $-0.04^{+0.01}_{-0.01}$ \\
        VCC2019 & $4.85^{+0.63}_{-0.56}$ & $-0.21^{+0.04}_{-0.04}$ & $-0.16^{+0.02}_{-0.02}$ & $8.46^{+1.3}_{-1.13}$ & $-0.55^{+0.06}_{-0.06}$ & $-0.18^{+0.02}_{-0.02}$ \\
        CIRCINUS & $0.16^{+0.04}_{-0.03}$ & $0.22^{+0.19}_{-0.19}$ & $0.60^{+0.12}_{-0.12}$ & $0.26^{+0.02}_{-0.02}$ & $0.12^{+0.16}_{-0.16}$ & $0.65^{+0.04}_{-0.04}$ \\
        ESO59-01 & $5.54^{+0.34}_{-0.32}$ & $-1.74^{+0.03}_{-0.03}$ & $0.16^{+0.08}_{-0.08}$ & $8.92^{+1.02}_{-0.92}$ & $-1.68^{+0.06}_{-0.06}$ & $0.36^{+0.07}_{-0.07}$ \\
        FCC119 & $0.2^{+0.03}_{-0.03}$ & $-0.43^{+0.09}_{-0.09}$ & $-0.22^{+0.49}_{-0.49}$ & $2.05^{+0.17}_{-0.16}$ & $-0.42^{+0.02}_{-0.02}$ & $-0.06^{+0.02}_{-0.02}$ \\
        FCC148 & $1.83^{+0.07}_{-0.07}$ & $0.31^{+0.01}_{-0.01}$ & $-0.1^{+0.03}_{-0.03}$ & $2.52^{+0.16}_{-0.15}$ & $-0.05^{+0.01}_{-0.01}$ & $0.06^{+0.08}_{-0.08}$ \\
        
        \hline
	\end{tabular}}
\end{table*}

\begin{table*}
	\label{tab:ppxf result3}
        \hspace{-1.5cm}
        \renewcommand\arraystretch{1.5}
        \setlength{\tabcolsep}{3mm}{
	\begin{tabular}{ccccccc}
        \hline
        $\rm Gal$ & $ Age_{\rm NSC} $ (Gyr) & $[\rm M/\rm H]_{\rm NSC}$ & $[\alpha/\rm Fe]_{\rm NSC}$ & $Age_{\rm host}$ (Gyr) & $[\rm M/\rm H]_{\rm host}$ & $[\alpha/\rm Fe]_{\rm host}$
        \\
        \hline
        FCC153 & $3.92^{+0.22}_{-0.21}$ & $0.31^{+0.03}_{-0.03}$ & $-0.07^{+0.04}_{-0.04}$ & $4.36^{+0.17}_{-0.17}$ & $0.2^{+0.02}_{-0.02}$ & $0.08^{+0.01}_{-0.01}$ \\
        FCC170 & $12.0^{+0.17}_{-0.17}$ & $0.19^{+0.02}_{-0.02}$ & $0.2^{+0.02}_{-0.02}$ & $12.93^{+0.31}_{-0.31}$ & $-0.18^{+0.02}_{-0.02}$ & $0.22^{+0.01}_{-0.01}$ \\
        FCC177 & $1.57^{+0.04}_{-0.04}$ & $0.33^{+0.01}_{-0.01}$ & $0.37^{+0.0}_{-0.0}$ & $4.45^{+0.12}_{-0.12}$ & $0.04^{+0.01}_{-0.01}$ & $0.05^{+0.0}_{-0.0}$ \\
        FCC182 & $13.63^{+0.15}_{-0.15}$ & $0.09^{+0.01}_{-0.01}$ & $0.05^{+0.01}_{-0.01}$ & $5.83^{+0.37}_{-0.35}$ & $-0.29^{+0.02}_{-0.02}$ & $0.11^{+0.01}_{-0.01}$ \\
        FCC188 & $5.99^{+0.21}_{-0.2}$ & $-1.01^{+0.02}_{-0.02}$ & $0.07^{+0.03}_{-0.03}$ & $5.3^{+0.77}_{-0.67}$ & $-0.83^{+0.09}_{-0.09}$ & $0.23^{+0.1}_{-0.1}$ \\
        FCC190 & $4.88^{+0.15}_{-0.14}$ & $0.29^{+0.02}_{-0.02}$ & $-0.02^{+0.06}_{-0.06}$ & $10.36^{+0.77}_{-0.72}$ & $-0.26^{+0.03}_{-0.03}$ & $0.14^{+0.01}_{-0.01}$ \\
        FCC193 & $13.74^{+0.02}_{-0.02}$ & $0.39^{+0.0}_{-0.0}$ & $0.13^{+0.02}_{-0.02}$ & $10.08^{+0.28}_{-0.27}$ & $-0.08^{+0.02}_{-0.02}$ & $0.15^{+0.01}_{-0.01}$ \\
        FCC202 & $8.27^{+0.51}_{-0.48}$ & $-0.85^{+0.04}_{-0.04}$ & $-0.16^{+0.03}_{-0.03}$ & $5.08^{+0.33}_{-0.31}$ & $-0.35^{+0.03}_{-0.03}$ & $0.14^{+0.0}_{-0.0}$ \\
        FCC207 & $0.61^{+0.15}_{-0.12}$ & $0.23^{+0.08}_{-0.08}$ & $0.01^{+3.77}_{-3.77}$ & $10.61^{+1.33}_{-1.18}$ & $-0.61^{+0.1}_{-0.1}$ & $-0.14^{+0.02}_{-0.02}$ \\
        FCC211 & $3.79^{+0.67}_{-0.57}$ & $-1.03^{+0.05}_{-0.05}$ & $0.67^{+0.05}_{-0.05}$ & $4.09^{+0.33}_{-0.3}$ & $-0.81^{+0.04}_{-0.04}$ & $-0.04^{+0.01}_{-0.01}$ \\
        FCC215 & $5.28^{+0.53}_{-0.49}$ & $-1.57^{+0.06}_{-0.06}$ & $0.57^{+0.05}_{-0.05}$ & $1.41^{+0.31}_{-0.25}$ & $-1.5^{+0.24}_{-0.24}$ & $-0.23^{+2.06}_{-2.06}$ \\
        FCC222 & $5.37^{+0.82}_{-0.71}$ & $-1.67^{+0.07}_{-0.07}$ & $-0.29^{+0.53}_{-0.53}$ & $7.5^{+0.59}_{-0.55}$ & $-0.64^{+0.04}_{-0.04}$ & $0.11^{+0.01}_{-0.01}$ \\
        FCC223 & $7.07^{+0.79}_{-0.71}$ & $-1.95^{+0.07}_{-0.07}$ & - & $9.9^{+1.33}_{-1.17}$ & $-1.4^{+0.08}_{-0.08}$ & - \\
        FCC227 & $4.59^{+6.41}_{-2.67}$ & $-1.15^{+0.67}_{-0.67}$ & - & $1.55^{+1.93}_{-0.86}$ & $-0.74^{+0.72}_{-0.72}$ & - \\
        FCC245 & $8.48^{+2.89}_{-2.15}$ & $-0.84^{+0.26}_{-0.26}$ & - & $10.06^{+1.85}_{-1.57}$ & $-0.86^{+0.15}_{-0.15}$ & - \\
        FCC249 & $10.63^{+0.19}_{-0.19}$ & $0.29^{+0.03}_{-0.03}$ & $0.31^{+0.01}_{-0.01}$ & $12.67^{+0.39}_{-0.37}$ & $-0.47^{+0.02}_{-0.02}$ & $0.37^{+0.01}_{-0.01}$ \\
        FCC255 & $1.3^{+0.11}_{-0.1}$ & $0.06^{+0.03}_{-0.03}$ & $-0.18^{+0.07}_{-0.07}$ & $3.96^{+0.18}_{-0.17}$ & $-0.27^{+0.01}_{-0.01}$ & $0.09^{+0.01}_{-0.01}$ \\
        FCC277 & $3.18^{+0.16}_{-0.15}$ & $-0.19^{+0.04}_{-0.04}$ & $0.31^{+0.03}_{-0.03}$ & $8.1^{+0.71}_{-0.65}$ & $-0.78^{+0.03}_{-0.03}$ & $0.67^{+0.01}_{-0.01}$ \\
        FCC301 & $3.4^{+0.19}_{-0.18}$ & $0.17^{+0.04}_{-0.04}$ & $-0.0^{+0.01}_{-0.01}$ & $7.96^{+0.35}_{-0.34}$ & $-0.72^{+0.03}_{-0.03}$ & $0.16^{+0.01}_{-0.01}$ \\
        FCC306 & $0.25^{+0.05}_{-0.04}$ & $-0.88^{+0.23}_{-0.23}$ & $0.35^{+0.37}_{-0.37}$ & $0.47^{+0.05}_{-0.04}$ & $-0.99^{+0.05}_{-0.05}$ & $0.1^{+0.07}_{-0.07}$ \\
        FCC310 & $9.7^{+0.24}_{-0.24}$ & $0.3^{+0.01}_{-0.01}$ & $0.0^{+0.01}_{-0.01}$ & $6.26^{+0.2}_{-0.19}$ & $-0.22^{+0.01}_{-0.01}$ & $0.08^{+0.0}_{-0.0}$ \\
        FCC47 & $12.72^{+0.36}_{-0.35}$ & $-0.3^{+0.03}_{-0.03}$ & $0.46^{+0.01}_{-0.01}$ & $10.77^{+0.4}_{-0.39}$ & $-0.67^{+0.02}_{-0.02}$ & $0.31^{+0.05}_{-0.05}$ \\
        FCCB1237 & $5.39^{+2.22}_{-1.57}$ & $-0.7^{+0.26}_{-0.26}$ & - & $10.39^{+1.63}_{-1.41}$ & $-0.12^{+0.11}_{-0.11}$ & - \\
        IC1959 & $11.55^{+1.34}_{-1.2}$ & $-0.83^{+0.07}_{-0.07}$ & $-0.07^{+0.03}_{-0.03}$ & $1.61^{+0.2}_{-0.18}$ & $-1.43^{+0.05}_{-0.05}$ & $0.41^{+0.09}_{-0.09}$ \\
        IC5332 & $5.33^{+0.48}_{-0.44}$ & $-0.59^{+0.03}_{-0.03}$ & $-0.4^{+0.14}_{-0.14}$ & $5.33^{+0.48}_{-0.44}$ & $-0.59^{+0.03}_{-0.03}$ & $0.09^{+0.04}_{-0.04}$ \\
        KK197 & $4.99^{+0.51}_{-0.46}$ & $-2.16^{+0.06}_{-0.06}$ & - & $4.85^{+0.93}_{-0.78}$ & $-1.32^{+0.11}_{-0.11}$ & - \\
        KKS58 & $0.64^{+0.11}_{-0.09}$ & $-1.87^{+0.09}_{-0.09}$ & $0.14^{+0.02}_{-0.02}$ & $0.14^{+0.04}_{-0.03}$ & $-1.84^{+0.2}_{-0.2}$ & $-0.07^{+0.02}_{-0.02}$ \\
        NGC1487 & $1.91^{+0.34}_{-0.29}$ & $-1.2^{+0.12}_{-0.12}$ & $-0.17^{+0.28}_{-0.28}$ & $0.97^{+0.11}_{-0.1}$ & $-0.59^{+0.05}_{-0.05}$ & $0.85^{+0.03}_{-0.03}$ \\
        
        \hline
	\end{tabular}}
\end{table*}

\begin{table*}
	\label{tab:ppxf result4}
        \hspace{-1.5cm}
        \renewcommand\arraystretch{1.5}
        \setlength{\tabcolsep}{3mm}{
	\begin{tabular}{ccccccc}
        \hline
        $\rm Gal$ & $ Age_{\rm NSC} $ (Gyr) & $[\rm M/\rm H]_{\rm NSC}$ & $[\alpha/\rm Fe]_{\rm NSC}$ & $Age_{\rm host}$ (Gyr) & $[\rm M/\rm H]_{\rm host}$ & $[\alpha/\rm Fe]_{\rm host}$
        \\
        \hline
        NGC1705 & $12.56^{+0.6}_{-0.57}$ & $-1.66^{+0.07}_{-0.07}$ & $-0.03^{+0.04}_{-0.04}$ & $0.21^{+0.02}_{-0.02}$ & $-1.53^{+0.06}_{-0.06}$ & $-0.32^{+0.11}_{-0.11}$ \\
        NGC1796 & $0.29^{+0.12}_{-0.09}$ & $-0.62^{+0.19}_{-0.19}$ & $0.72^{+0.01}_{-0.01}$ & $12.74^{+0.62}_{-0.59}$ & $-1.32^{+0.04}_{-0.04}$ & $0.23^{+0.0}_{-0.0}$ \\
        NGC2835 & $1.36^{+0.14}_{-0.13}$ & $-0.49^{+0.03}_{-0.03}$ & $0.39^{+0.06}_{-0.06}$ & $1.15^{+0.1}_{-0.09}$ & $-0.18^{+0.03}_{-0.03}$ & $0.11^{+0.03}_{-0.03}$ \\
        NGC3274 & $1.2^{+0.15}_{-0.13}$ & $-2.0^{+0.06}_{-0.06}$ & $-0.07^{+0.05}_{-0.05}$ & $2.1^{+0.37}_{-0.31}$ & $-1.24^{+0.08}_{-0.08}$ & $0.88^{+0.03}_{-0.03}$ \\
        NGC3368 & $1.53^{+0.09}_{-0.08}$ & $0.23^{+0.04}_{-0.04}$ & $0.0^{+0.04}_{-0.04}$ & $5.03^{+0.38}_{-0.35}$ & $-0.18^{+0.04}_{-0.04}$ & $0.13^{+0.01}_{-0.01}$ \\
        NGC3489 & $3.15^{+0.39}_{-0.35}$ & $0.15^{+0.08}_{-0.08}$ & $0.14^{+0.03}_{-0.03}$ & $3.1^{+0.22}_{-0.2}$ & $-0.23^{+0.04}_{-0.04}$ & $0.12^{+0.09}_{-0.09}$ \\
        NGC3593 & $5.33^{+0.8}_{-0.7}$ & $-0.28^{+0.15}_{-0.15}$ & $0.11^{+0.02}_{-0.02}$ & $12.07^{+0.86}_{-0.8}$ & $-0.49^{+0.03}_{-0.03}$ & $0.13^{+0.11}_{-0.11}$ \\
        NGC4592 & $0.13^{+0.02}_{-0.02}$ & $-0.22^{+0.07}_{-0.07}$ & $-0.02^{+0.08}_{-0.08}$ & $0.49^{+0.04}_{-0.04}$ & $-0.61^{+0.02}_{-0.02}$ & $0.11^{+0.06}_{-0.06}$ \\
        NGC5253 & $0.21^{+0.05}_{-0.04}$ & $-1.47^{+0.12}_{-0.12}$ & - & $0.54^{+0.1}_{-0.09}$ & $-1.67^{+0.1}_{-0.1}$ & - \\
        NGC853 & $4.07^{+0.41}_{-0.38}$ & $-1.42^{+0.05}_{-0.05}$ & $-0.0^{+0.0}_{-0.0}$ & $3.13^{+0.39}_{-0.34}$ & $-1.6^{+0.06}_{-0.06}$ & $0.36^{+0.06}_{-0.06}$ \\
        UGC3755 & $0.09^{+0.02}_{-0.02}$ & $-1.64^{+0.3}_{-0.3}$ & - & $1.82^{+0.24}_{-0.21}$ & $-1.37^{+0.12}_{-0.12}$ & - \\
        UGC5889 & $5.88^{+0.68}_{-0.61}$ & $-1.28^{+0.05}_{-0.05}$ & $0.45^{+0.07}_{-0.07}$ & $1.02^{+0.19}_{-0.16}$ & $-0.51^{+0.05}_{-0.05}$ & $-0.01^{+0.02}_{-0.02}$ \\
        UGC8041 & $0.35^{+0.05}_{-0.04}$ & $-0.51^{+0.07}_{-0.07}$ & $-0.12^{+0.0}_{-0.0}$ & $3.34^{+0.39}_{-0.35}$ & $-0.47^{+0.05}_{-0.05}$ & $0.48^{+0.27}_{-0.27}$ \\
        \hline
	\end{tabular}}
\end{table*}

\end{document}